\documentclass[11pt]{article}

\usepackage[final]{acl}
\usepackage{enumitem}

\usepackage{times}
\usepackage{latexsym}
\usepackage{amsmath}
\usepackage[T1]{fontenc}

\usepackage[most]{tcolorbox}
\usepackage{listings}

\tcbuselibrary{breakable,skins,listings}

\lstdefinestyle{promptstyle}{
  basicstyle=\ttfamily\small,
  breaklines=true,
  columns=fullflexible,
  keepspaces=true,
  showstringspaces=false
}

\newtcolorbox{promptbox}[2][]{
  enhanced,
  breakable,
  colback=white,
  colframe=black!60,
  boxrule=0.6pt,
  arc=1mm,
  title={#2},
  fonttitle=\bfseries,
  label={#1}
}

\usepackage[utf8]{inputenc}

\usepackage{microtype}

\usepackage{inconsolata}

\usepackage{booktabs,makecell,multirow,array}

\usepackage{booktabs}
\usepackage{threeparttable}
\usepackage{graphicx}
\usepackage{graphicx}

\usepackage{listings}
\usepackage[dvipsnames]{xcolor}
\colorlet{lightgray}{White!30!lightgray}
\colorlet{lightblue}{White!70!MidnightBlue}

\usepackage{xspace}

\usepackage[textsize=small,backgroundcolor=green,linecolor=ForestGreen]{todonotes}

\lstdefinelanguage{json}{
    basicstyle=\ttfamily\small,
    keywordstyle=\color{blue},
    stringstyle=\color{teal},
    showstringspaces=false,
    breaklines=true,
    frame=single
}

\usepackage{algorithm}
\usepackage{algpseudocode}

\usepackage{cleveref}
\crefname{figure}{Figure}{Figures}
\crefname{equation}{Equation}{Equations}
\crefname{appendix}{Appendix}{Appendices}
\crefname{table}{Table}{Tables}
\crefname{algorithm}{Algorithm}{Algorithms}
\crefname{section}{Section}{Sections}

\title{\ours: Simulating Individual Human Preference Judgments\\with Evaluator-Specific Demonstration Data}

\author{
 \textbf{Zeyu He\textsuperscript{1*}},
 \textbf{Xuan Qi\textsuperscript{2}},
 \textbf{Subramanian Chidambaram\textsuperscript{2}},
 \textbf{Zhichao Xu\textsuperscript{2}},
 \\
 \textbf{Vinayak Arannil\textsuperscript{2}},
 \textbf{Lydia Chilton\textsuperscript{2,3}},
  \textbf{Alex C. Williams\textsuperscript{2}}
\\
 \textsuperscript{1}Pennsylvania State University,
 \textsuperscript{2}AWS AI Fundamental Research,
 \textsuperscript{3}Columbia University
\\
 \small{
   \textbf{Correspondence:} \href{mailto:zeyuhe@psu.edu}{zeyuhe@psu.edu}
 }
}

\newcommand{\ours}{\textsc{PersonaJudge}\xspace}

\begin{document}
\maketitle

\def\thefootnote{*}\footnotetext{Work completed during ZH's Amazon internship.}
\def\thefootnote{\arabic{footnote}}

\begin{abstract}
Large language models increasingly serve as judges in AI evaluation, but current approaches rely on consensus preferences that ignore individual evaluator variation.
We propose a novel simulation approach that combines categorical judgments with evaluator-specific auxiliary data---retrospective reasoning traces and interface telemetry---to enable LLM-based simulation of individual evaluators via in-context learning. 
We conduct a systematic empirical study of this approach using multi-facet data from 32 trained annotators across 4,200 preference judgments in a $4\times4\times4$ factorial design. 
Our key findings: 
(1) The simulation approach achieves up to 9.9 percentage point improvements over the Base Judge;
(2) Reasoning traces provide the largest gains with higher collection efforts, while interface telemetry often hurts rather than helps performance despite being cheaper to collect.
(3) Simulation difficulty is systematic, predicted by an evaluator's neutral usage (most clearly on Helpfulness) and divergence from consensus; the neutral-usage tendency---rather than simulatability itself---is the cross-task-stable property ($r=0.728$).
These results establish both the potential and limits of evaluator-specific auxiliary data for personalized evaluation, offering methodological insights for scaling individual-aware AI assessment.
\end{abstract}

\section{Introduction}
\label{sec:intro}
\begin{figure}
    \centering
    \includegraphics[width=0.95\columnwidth]{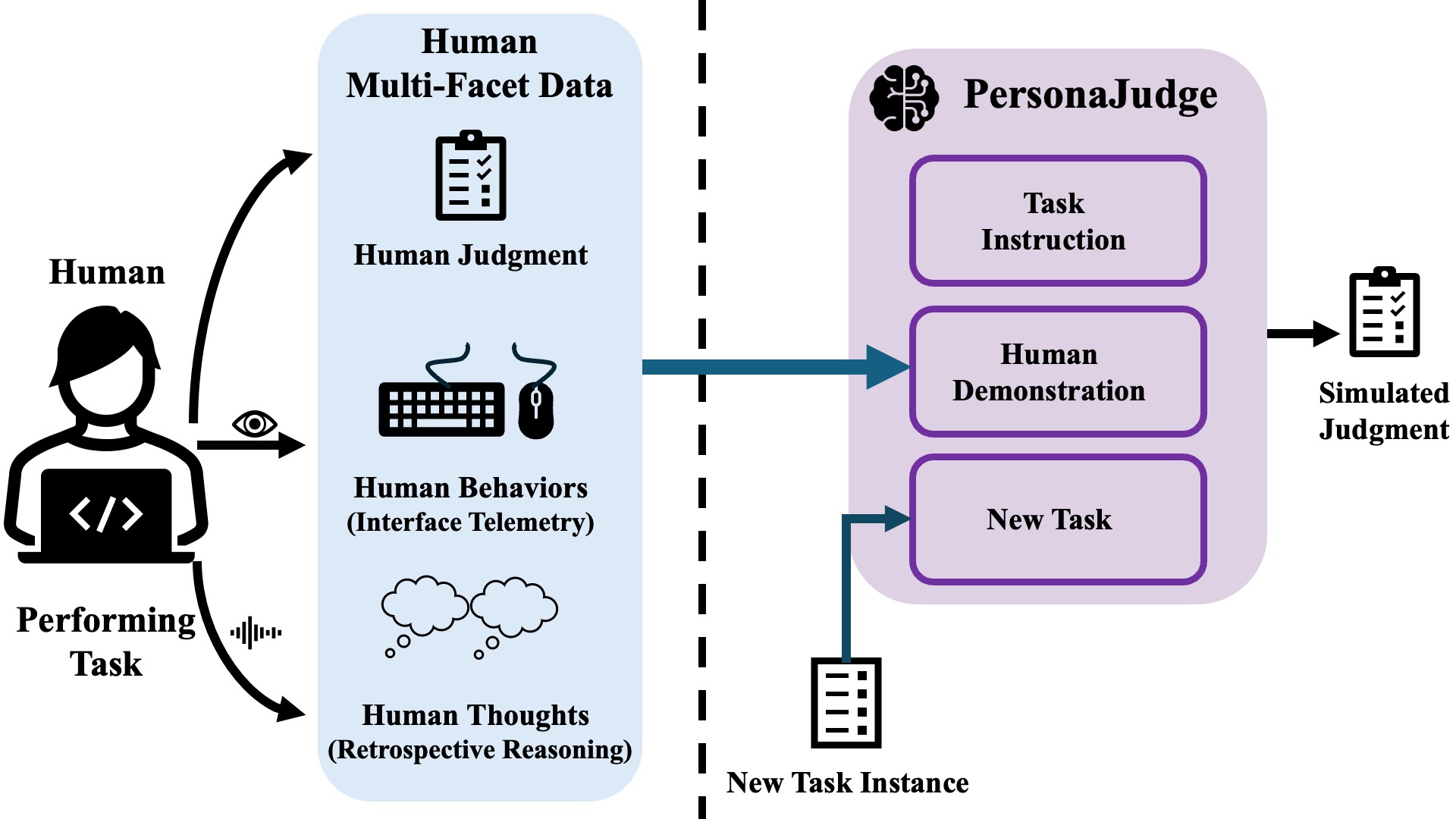}
    \caption{\textbf{\ours~workflow.} 
A human evaluator first completes judgment tasks, producing three complementary signals: a categorical judgment, interface telemetry, and retrospective reasoning. \ours~organizes these signals into evaluator-specific demonstrations and combines them with the task instruction and a new task instance. An LLM then uses this information to simulate the target evaluator's judgment.}
    \label{fig:personajudge-workflow}
\end{figure}
Large Language Models (LLMs) are increasingly used as judges for model comparison, reward modeling, and benchmark construction because they offer a scalable alternative to human evaluation~\cite{bai2022constitutional,liu-etal-2023-g,zheng2023judging, lambert2024rewardbench}.
However, most LLM-as-Judge pipelines target an aggregate signal: models are trained, aligned, or evaluated against pooled preferences that collapse across annotators~\cite{stiennon2020learning,ouyang2022training,bai2022training,zhang2024diverging}, so they simulate crowd-level consensus rather than any specific evaluator. This is a key limitation for open-ended evaluation, where a single shared standard may not exist and annotators apply different criteria to the same outputs~\cite{van2021human,davani2022dealing}. 
In such settings disagreement is often meaningful rather than noise~\cite{basile-etal-2021-need,fleisig2023majority,frenda2025perspectivist}. 
Even LLM judges that match average ratings can miss group- and person-level differences~\cite{movva2024annotation}.
Current models struggle to infer individual preferences from value-laden statements~\cite{jiang-etal-2025-language}. 
Matching consensus thus does not guarantee faithful simulation of any individual evaluator.

This motivates a different target: individual evaluator simulation, which can support evaluation auditing, per-person reward modeling, and fairness analysis of whose perspectives evaluation pipelines underrepresent.
Prior personalized methods condition on persona descriptions or compact preference profiles~\cite{dong-etal-2024-llm,wang2024learning}, but represent people through summaries rather than their own decision traces.
We instead ask: \emph{given an evaluator's historical judgments and decision traces, can an LLM simulate how that evaluator will judge a new instance?}

We argue that this individual evaluator simulation requires richer supervision than final labels alone. A final label captures an outcome, but not the process that produced it. 
We therefore propose \ours~(\cref{fig:personajudge-workflow}), an evaluator-specific framework that uses multi-facet evaluator data: (1) a primary \textbf{categorical judgment}, (2) \textbf{interface telemetry} capturing how evaluators inspect the task, and (3) \textbf{retrospective reasoning} giving post-hoc explanations of why decisions were made.

Using this framework, we evaluate whether evaluator-specific data can improve simulation over the Base Judge, which complementary signals contribute most, and how performance varies across evaluators. We conduct a systematic empirical study with data from 32 trained annotators and 4,200 preference judgments in a factorial design. Extensive experiments show that \ours improves simulation accuracy over the Base Judge by up to 9.9 points; retrospective reasoning gives the largest gains, while interface telemetry often reduces performance. 
We also find that simulation difficulty is systematic---predicted by an evaluator's neutral usage and divergence from consensus---with neutral usage a stable cross-task trait.

In summary, our contributions are:
\begin{itemize}[leftmargin=*]
    \item We propose a multi-facet simulation framework \ours for evaluator-specific judgment behavior using categorical judgments, interface telemetry, and retrospective reasoning.
    \item We conduct a systematic empirical analysis of how complementary evaluator-specific signals, model choice, and demonstration count affect the simulation performance.
    \item We characterize evaluator-level simulatability, showing that simulation difficulty varies widely but systematically across evaluators---predicted by their neutral usage and divergence from consensus---and that neutral usage is a stable, cross-task trait.
\end{itemize}

\section{Related Work}
\label{sec:related_work}
We review three relevant lines of research: LLM-as-Judge, individualized alignment, and process-oriented behavioral/cognitive traces for modeling human decision behavior.
\paragraph{Language models as judges}
LLMs are increasingly used as evaluators to simulate human judgments.
LLM-as-a-Judge methods have shown strong alignment with aggregate human judgments, including improved correlations with human ratings and high agreement with human preferences~\cite{liu-etal-2023-g,zheng2023judging} and have rapidly become a standard evaluation paradigm~\cite{gu2026survey,li2024llms}, but they rely on standardized rubrics that obscure individual variation in judgment styles and criteria application.
Studies have highlighted concerns about the biases and robustness in these methods~\cite{wang2023large, zheng2023judging, shi2024judging,ye2024justiceprejudicequantifyingbiases}. 
Moreover, the reproducibility of LLM-based evaluation has been questioned because model behavior can drift across updates and evaluation criteria may evolve during the validation process~\cite{chen2024chatgpt,shankar2024validates}.

Traditional crowdsourcing approaches similarly aggregate multiple annotations to improve reliability, but such aggregation can obscure meaningful differences in annotator judgments and consistency~\cite{snow2008cheap,williams2017deja}. 
Popular preference datasets and pipelines are often formulated around binary pairwise comparisons~\cite{bai2022training, stiennon2020learning, nakano2021webgpt,pmlr-v162-ethayarajh22a}, which can make it difficult to represent neutral, uncertain, or mixed judgments that arise in realistic scenarios involving underspecified criteria. 

\paragraph{Individualized alignment}
Beyond using LLMs as judges during evaluation, a related line of work examines how human preferences are modeled in alignment pipelines. 
These approaches have improved models' ability to align with human feedback, but they largely treat preferences as outcome labels and rarely model how individual evaluators arrive at their decisions.
Reinforcement Learning from Human Feedback approaches \cite{bai2022training,ouyang2022training} train reward models on aggregated human preference data
using techniques like Bradley-Terry modeling~\cite{bradley1952rank}.
Later works such as Constitutional AI \cite{bai2022constitutional} and RLAIF \cite{lee2023rlaif} extend this paradigm by incorporating AI feedback, but they aim to learn a single, static preference function.
Recent multi-objective and diversity-aware alignment methods model heterogeneous preferences through Pareto trade-offs or mixtures of reward models, but they remain formulated at the population level rather than simulating a particular evaluator~\cite{rame2023rewarded,pmlr-v235-chakraborty24b}.
Persona-based conditioning can be unreliable, because available persona descriptions can be too sparse or simplistic to predict task-specific preferences~\cite{dong-etal-2024-llm}.
Even when users attempt to guide an LLM more directly, they may struggle to translate their implicit standards into explicit labeling instructions: iterative prompt refinement without gold labels does not reliably improve labeling accuracy~\cite{he2025prompting}.
Other work models or simulates individual responses by conditioning on self-reports---interviews and surveys~\cite{park2026llm} or persona and demographics~\cite{hu2025inews}---but relies on elicited descriptions of a person rather than an evaluator's own decision traces for preference judgment.
These limitations suggest that evaluator-specific simulation requires richer supervision beyond preference outcomes, including behavioral traces of how evaluators inspect evidence and explanatory traces of how they justify their decisions. 

\paragraph{Behavioral and cognitive traces}
Prior work on process-oriented supervision points to two signals that may complement outcome-only preference modeling.

Behavior-wise, recent work has begun to model people through their interaction traces rather than only their final outputs. 
For example, \citet{shaikh2025creating} collect computer-use observations to build general user models from everyday interactions, showing that behavioral traces can support inference about users' preferences.
\citet{wang2025opera}'s OPeRA dataset goes further by collecting browser observations, fine-grained web actions, and self-reported rationales to benchmark personalized human behavior simulation in online shopping.
These studies suggest that interaction traces can expose how users gather evidence and act in context, but they focus on user modeling or next-action prediction rather than open-ended judgment simulation.
Most closely, \citet{de2025tracing} augment preference judgments with annotators' mouse-tracking reading traces, but analyze the dataset rather than simulating annotators, and use study-time traces rather than deployment telemetry with retrospective reasoning.

Cognitive-wise, a separate line of work shows the value of natural language explanations and feedback for making human reasoning more legible to learning systems.
Chain-of-Thought prompting and related evaluation methods show that intermediate reasoning can improve model performance and judgment quality~\cite{wei2022chain, kojima2022large, liu-etal-2023-g}, but they rely on model-generated reasoning rather than human-authored judgment explanations.
More broadly, prior work shows that human-provided explanations and feedback can support model development~\cite{stumpf2007toward, wiegreffe2021teach}.
However, these efforts are not aimed at simulating how a particular evaluator makes subjective judgments in open-ended comparison settings.

Existing work has explored richer evaluator-specific signals, but has not examined whether behavioral and cognitive traces jointly support evaluator-specific judgment simulation.
Our work addresses this gap using interface telemetry and retrospective reasoning as complementary supervision.

\section{Simulating Individual Preference Judgments with \ours}
\label{sec:approach}
We introduce \ours, a novel approach for simulating individuals' preference judgments using LLMs and their ability to perform in-context learning (ICL)~\cite{dong2022survey} with multi-facet, evaluator-specific demonstration data. 
Unlike consensus-based preference modeling \cite{ouyang2022training, bai2022training}, \ours~simulates individual evaluators' decision-making processes by using a primary categorical judgment signal, interface telemetry, and retrospective reasoning.
\subsection{Problem Formulation}

\paragraph{Setup}
Given a target evaluator $e_j$ and a target evaluation instance $x=(c,r^A,r^B)$,
where $c$ is the conversation context and $r^A,r^B$ are the two candidate responses,
\ours~simulates $e_j$'s three-class judgment $y\in\{A,N,B\}$ (prefer A, neutral,
prefer B) through in-context learning over evaluator-specific demonstrations. Let
$x_i=(c_i,r_i^A,r_i^B)$ denote a prior instance previously judged by $e_j$ with label
$y_{i,j}\in\{A,N,B\}$ and an optional explanatory signal $s_{i,j}$ (interface telemetry
and/or retrospective reasoning).

\paragraph{Two-round elicitation}
We elicit the judgment as two binary decisions. A \emph{preference-detection} step first
determines whether $e_j$ expresses a preference; a \emph{direction} step, invoked only
when a preference is detected, then determines which option is preferred. 
Define preference label $b(y)=\textsc{neutral}$ if $y=N$ and $b(y)=\textsc{pref}$ otherwise, and the direction $\delta(y)=y$ for $y\in\{A,B\}$. 
Both rounds use the same prompt structure---an instruction header, $k$ evaluator-specific demonstrations, and the
unlabeled query $\mathrm{Q}(x)$---but differ in their round-specific label mapping and sampling constraints: Round 1 maps labels to preference vs.\ no preference, whereas Round 2 uses directional preference labels only (Appendix~\ref{sim-prompt}).
\paragraph{Round 1 (preference detection)} \ours~constructs the prompt
{\small\begin{equation}
\mathrm{P}^{1}(e_j,x)=\mathrm{Intro}^{1}\oplus\bigoplus_{i=1}^{k}
   \mathrm{Demo}\!\big(x_i,\,b(y_{i,j}),\,s_{i,j}\big)\oplus \mathrm{Q}(x), \notag
\end{equation}}
where each demonstration carries the binary label
$b(y_{i,j})\in\{\textsc{neutral},\textsc{pref}\}$. The model returns
$\hat{p}\in\{\textsc{neutral},\textsc{pref}\}$.

\paragraph{Round 2 (direction)} If $\hat{p}=\textsc{pref}$, \ours~constructs
{\small\begin{equation}
\mathrm{P}^{2}(e_j,x)=\mathrm{Intro}^{2}\oplus
   \!\!\bigoplus_{i:\,y_{i,j}\neq N}\!\!
   \mathrm{Demo}\!\big(x_i,\,\delta(y_{i,j}),\,s_{i,j}\big)\oplus \mathrm{Q}(x), \notag
\end{equation}}
using demonstrations sampled from the same evaluator-specific pool, restricted to
instances in which $e_j$ expressed a directional preference, each labeled with its
direction $\delta(y_{i,j})\in\{A,B\}$. The model returns $\hat{d}\in\{A,B\}$.

\paragraph{Composition} The simulated judgment recovers the full three-class space,
\begin{equation}
\hat{y}=
\begin{cases}
N        & \text{if } \hat{p}=\textsc{neutral},\\
\hat{d}  & \text{if } \hat{p}=\textsc{pref}, \notag
\end{cases}
\end{equation}
making the neutral decision explicit while keeping each round a binary choice. In the
prompts, the two outcomes of each round are encoded as the rating numbers $1$ and $2$
(Appendix \ref{sim-prompt}).

\subsection{Demonstration Data Model}
\label{subsec:framework}

\ours~demonstrations combine categorical preference labels \texttt{\{Prefer A, Neutral, Prefer B\}} with two types of explanatory data: \textbf{interface telemetry} and \textbf{retrospective reasoning}.
These two signals complement categorical judgments in distinct ways:
Interface telemetry provides an implicit trace of how evaluators inspect tasks, while retrospective reasoning provides an explicit account of the criteria and trade-offs behind their decisions.
Unlike binary preference datasets~\cite{stiennon2020learning,bai2022training}, \ours~keeps \texttt{Neutral} as a valid third label instead of collapsing or removing neutral judgments.
This three-class judgment is the canonical stored demonstration label; round-specific binary labels are derived from it only at prompt construction time.


\subsubsection{Interface Telemetry Data} 

Interface telemetry augments categorical judgments with implicit behavioral traces of how evaluators inspect information during decision making---what was inspected, in what order, for how long, and whether regions were revisited~\cite{jasper2002mousetrace,franco2011applying}.
Process-tracing research shows that decision processes leave informative behavioral traces~\cite{payne1993adaptive,ford1989process}, and interaction data such as clicks and dwell time can help infer user goals and information access behavior~\cite{guo2010ready,liu2010understanding}.
Incorporating these traces lets \ours~condition on not only \emph{what} an evaluator decided, but also \emph{how} they inspected the task before deciding.

\subsubsection{Retrospective Reasoning Data}

Retrospective verbal reports can complement categorical judgments by providing explicit accounts of the evaluator's decision process~\cite{schulte2011role}: a final label alone does not reveal which aspects of the candidate responses an evaluator prioritized, nor why they selected \texttt{Neutral} when responses satisfied different criteria.
Such natural-language explanations can make decision-relevant considerations more explicit in a form language models can directly use~\cite{camburu2018snli,zhou2020towards}, surfacing evaluator-specific judgment criteria that labels alone do not.
Because these reports are collected after task completion, they avoid interfering with the judgment itself~\cite{ericsson1993protocol,vansomeren1994think}.
In our setting, this signal is especially useful for ICL-based simulation because it provides evaluator-specific judgment criteria beyond labels alone.
We operationalize this signal as post-task think-aloud transcripts.


\section{Study Design}
\label{sec:study_design}

We conduct an empirical study to evaluate the effectiveness of \ours~in simulating individual human judgments to address the research questions (RQs) introduced in \cref{sec:findings}. 



\subsection{Evaluation Task Setup}
\label{subsec:dataset}

We sample instances from Anthropic's Helpful and Harmless (HH) dataset~\cite{bai2022training}, a widely adopted, publicly available preference benchmark whose two distinct evaluation criteria---helpfulness and harmlessness---engage different evaluator reasoning and enable our cross-task analysis. 
We select 700 conversations each for the helpfulness and harmlessness evaluation tasks.
We recruited 32 trained annotators, 10 of whom contributed to both datasets, resulting in 21 evaluators per dataset.
Each evaluator completed 100 judgments per dataset, yielding 2,100 judgments for helpfulness and 2,100 for harmlessness.
We randomize the presentation order and implement position-bias controls by varying the response ordering across evaluators~\cite{wang2023large, zheng2023judging}.

\subsection{Multi-Facet Data Collection}
\label{sec:two-stage-workflow}
Since \ours~simulates individual human judgments by using evaluator-specific, multi-facet demonstrations, we obtain those demonstrations through a two-stage annotation workflow designed to preserve natural judgments while capturing complementary explanatory signals. 
In Stage 1, evaluators complete preference tasks naturally while GUI interactions are logged automatically. 
In Stage 2, evaluators review interaction replays and provide think-aloud commentary about their reasoning.
Full data collection details are provided in Appendix~\ref{app:data-collection}.


\paragraph{Task Interface}
We implement a browser-based interface with conversational contexts and paired responses. Evaluators select among \texttt{\{Prefer A, Neutral, Prefer B\}} with neutrality as an equally valid choice. Progressive disclosure reveals information as evaluators complete milestones, with all interactions automatically timestamped using a structured event taxonomy (See \cref{sec:implementation}).

\paragraph{Data Post-Processing}
After multi-facet data collection, we transform raw interface interaction logs and retrospective reasoning transcripts into the structured demonstration format used by \ours.
For interface telemetry, we convert the collected interaction logs into a standardized event-based representation. Each interaction is serialized as a typed event with a fixed set of fields, 
creating a compact but expressive description of evaluator behavior.
The full post-processed telemetry scheme used in ICL is provided in \cref{tab:telemetry}.
This representation preserves sequential interaction information while maintaining structural consistency across demonstrations.
For retrospective reasoning, we organize each think-aloud script into an ordered list of section-specific reasoning entries, where each entry contains a reasoning section and its associated content (\cref{fig:think-aloud}).



\subsection{Experimental Factors}
\label{subsec:factorial_design}

We employ a $4\times4\times4$ factorial design in which we generate simulations for three controlled factors:

\begin{itemize}[nosep,leftmargin=*]
\item \textbf{Demonstration Type}: We produce simulations using the four prompt-based configurations made possible by \ours: (1) J (judgment-only); (2) J+IT (judgments with interface telemetry); (3) J+RR (judgments with retrospective reasoning); and (4) J+IT+RR (judgments with interface telemetry and retrospective reasoning data).
\item \textbf{Demonstration Count}: We conduct simulations using a predetermined number of randomly-sampled demonstrations: (1) one demonstration, (2) two demonstrations, (3) four demonstrations, or (4) eight demonstrations.
\item \textbf{Simulation Model}: We conduct simulations using four performant LLMs: (1) Claude-3.5-Sonnet-V2, (2) Claude-3.7-Sonnet-V1, (3) DeepSeek-R1, and (4) Amazon Nova Premier.
\end{itemize}


\subsection{Experimental Setup}\label{subsec:experimental_setup}

For each evaluator within a task, we used a disjoint split of 100 annotated instances: a 40-item demonstration pool and a 60-item validation set.
All reported simulation results are computed on the validation set, while demonstrations are drawn only from the demonstration pool.
The evaluation items are never reused as in-context demonstrations.

For each human judgment in the validation set, we generated 64 simulated judgments by crossing three experimental factors mentioned in \cref{subsec:factorial_design}.
Each simulated judgment follows the two-round cascade (Algorithm~\ref{alg:personajudge}, Appendix~\ref{app:pseudocode}): a preference-detection round is run first, and a direction round is run only when the first round predicts a preference.
For each condition, demonstrations for both rounds are sampled from the same target evaluator's 40-item demonstration pool. Round 1 samples $k$ demonstrations balanced between \texttt{Neutral} and directional preference labels, while Round 2 samples $k$ demonstrations balanced between \texttt{Prefer A} and \texttt{Prefer B}. Exact balancing is not possible in the 1-shot setting.
Two helpfulness evaluators with only a few neutral labels are handled as a direction-only special case (Appendix~\ref{app:simulation-prompt}).
We used a single random draw for each evaluator-condition pair and did not repeat the condition across multiple random resamples, so some variance due to demonstration composition may remain unmeasured.

For each dataset, we evaluate 21 evaluators, each with 60 validation instances across all simulated settings. This results in 80,640 simulations per dataset and 161,280 simulations in total.

\subsection{Metrics and Baselines}
\label{subsec:metrics_baselines}
We report three-class accuracy over \texttt{\{Prefer A, Neutral, Prefer B\}}
by comparing each simulated judgment---composed from the two-round cascade---with the corresponding evaluator's original human judgment on the same validation instance. 
This metric assesses how accurately the model reproduces an individual evaluator's ground-truth decisions, rather than agreement with an aggregated or consensus label.
We preserve the full judgment space instead of collapsing evaluations into binary preferences as in prior works~\cite{bai2022training, stiennon2020learning}.

\paragraph{Primary baselines}
We use \textbf{Random Selection}, which chooses uniformly among the three labels (expected accuracy
$1/3$). The \textbf{Base Judge} (Model Judgment Baseline) is a zero-shot LLM-as-Judge
that produces a judgment with no evaluator-specific demonstrations and no persona,
using the same two-round protocol.
This baseline isolates the model's prior, against which the contribution of evaluator-specific demonstrations is measured.

\paragraph{Controls and reference predictors}
To interpret \emph{how} \ours~captures individual variation
(\cref{subsec:teachability_patterns}), we use one control and three reference
predictors, again scored against each evaluator's own labels.
(i) The \textbf{cross-evaluator control} holds the model, demonstration
type, and shot count fixed but draws demonstrations from \emph{non-target}
evaluators, averaging over all available non-target evaluators of each item, isolating the
benefit of evaluator-matched demonstrations from that of demonstrations in general.
(ii) The \textbf{oracle consensus} baseline predicts, for every evaluator,
an item's majority label among its three human annotations; it is an ``oracle''
because it uses held-out human labels the model never observes, and serves as the
strongest possible group prior.
(iii) The \textbf{global majority-class} baseline always predicts the single most frequent label across the entire task, pooled over all evaluators (e.g., \texttt{Neutral} on Harmlessness).
(iv) The \textbf{per-evaluator majority-class} baseline predicts each
individual's own most frequent label---the strongest label-frequency predictor available per person. 

\section{Results and Analysis}
\label{sec:findings}
Our empirical study addresses three questions:
\begin{itemize}[leftmargin=0pt, itemsep=0pt]
\item[] \textbf{[RQ1]}: Can evaluator-specific demonstrations improve over the Base Judge when simulating a specific evaluator?
\item[] \textbf{[RQ2]}: How do complementary evaluator-specific signals affect simulation performance, and how do these effects vary with model choice and demonstration count?
\item[] \textbf{[RQ3]}: How does simulation fidelity vary across evaluators, and which evaluator-level properties explain that variation and transfer across tasks?


\end{itemize}

\subsection{Main Results}
\subsubsection{RQ1: Overall Effectiveness}
\label{subsec:alignment_effectiveness}

\begin{figure}
    \centering
    \includegraphics[width=0.98\linewidth]{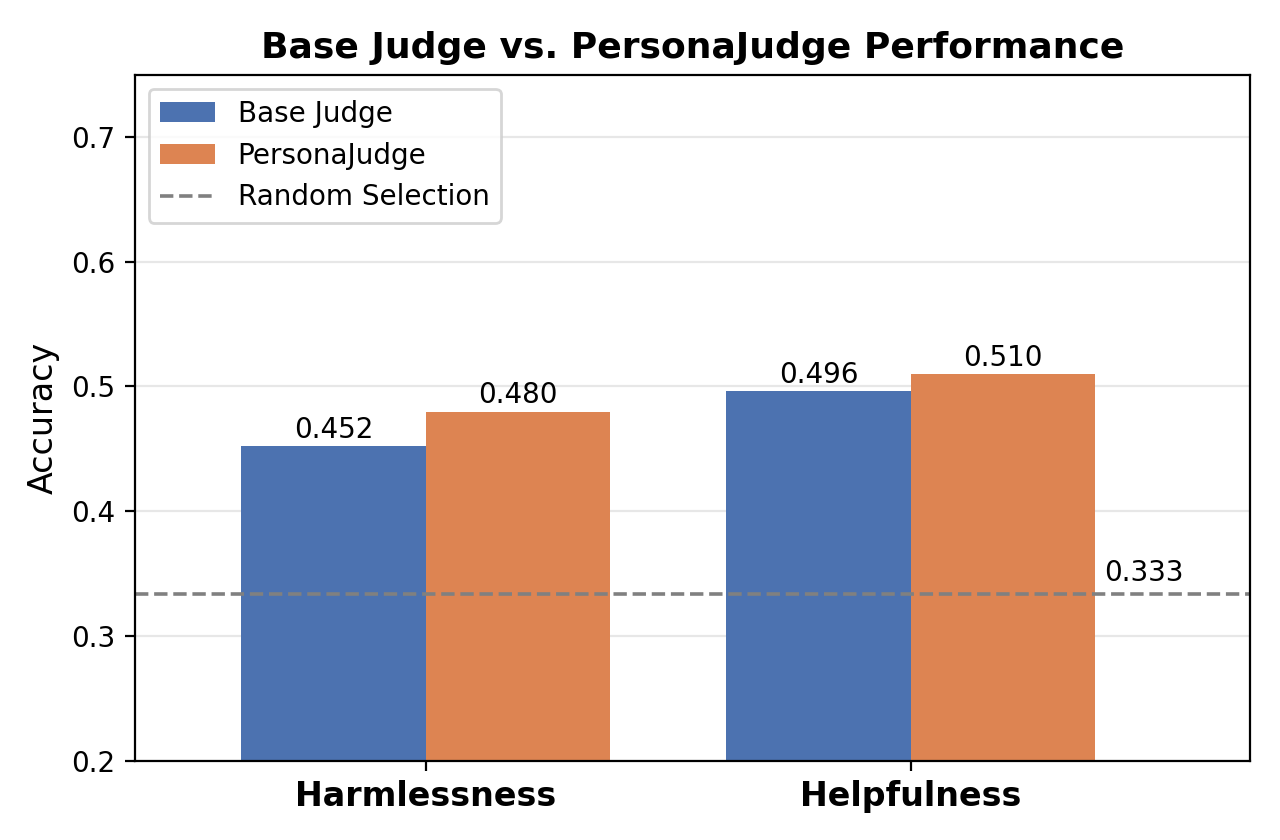}
    \caption{Accuracy comparison across random selection, the Base Judge without evaluator-specific demonstrations, and \textbf{\ours}. Detailed results in Appendix~\ref{app:sim-detail-result}.}
    \label{tab:main-acc}
\end{figure}

\Cref{tab:main-acc} presents our core results. 
We compare Random Selection, the Base Judge, and \ours, which conditions on evaluator-specific demonstrations.
\ours~results are averaged over the 64 simulation configurations defined by 4 demonstration types, 4 demonstration counts, and 4 models for each dataset.


The Base Judge outperforms random selection ($0.333$), reaching $0.452$ on
Harmlessness and $0.496$ on Helpfulness on average. 
\ours~improves on these non-personalized judges, reaching $0.480$ on Harmlessness and $0.510$ on Helpfulness ($+2.8$ and $+1.4$ points on average).

These averages are deliberately conservative: they pool all 64 model\,$\times$\,demonstration-type\,$\times$\,shot combinations, including suboptimal combinations.
Comparing each combination against its \emph{own model's} Base Judge, $43$ of $64$ combinations improve on Harmlessness and $50$ of $64$ on Helpfulness, and the \textbf{J+RR} family improves in $14/16$ and $16/16$ combinations, respectively. 
The combinations that fail to beat their Base Judge are concentrated in the \textbf{IT} conditions ($16$ of the $21$ Harmlessness misses and $9$ of the $14$ Helpfulness misses involve telemetry), foreshadowing the negative telemetry effect in \cref{subsec:data_contribution}. 

To verify that these gains reflect personalization rather than demonstrations in general, we compare \ours~against a cross-evaluator control that uses demonstrations from a non-target evaluator while holding the demonstration type, model, and shot count fixed.
\ours~outperforms this control by $+0.028$ on Harmlessness ($0.477$ vs.\ $0.450$, $p=0.019$) and $+0.044$ on Helpfulness ($0.515$ vs.\ $0.471$, $p<0.001$).\footnote{The control requires non-target evaluators' predictions, which exist only for items judged by $\ge2$ evaluators, so both \ours~and the control are evaluated on this matched subset ($84.1\%$ of pairs); \ours's accuracy on it ($0.477$/$0.515$) differs from the full-set average in \cref{tab:main-acc} ($0.480$/$0.510$). Full significance tests are in Appendix~\ref{app:significance}.}
The advantage is positive for most evaluators and becomes strongest with 4--8 demonstrations, suggesting that evaluator-specific examples provide genuine personalization signal rather than merely additional labeled context.

\subsubsection{RQ2: Signal and Factor Patterns}
\label{subsec:data_contribution}
\Cref{tab:demo_factors} reports marginal average simulation accuracy for each experimental factor.
Each block isolates one factor while averaging over the other two: \textbf{Demonstration Type} averages over all models and counts, \textbf{Model} averages over all demonstration types and counts, and \textbf{Demonstration Count} averages over all models and demonstration types.



\begin{table}[t]
\centering
\resizebox{0.9\columnwidth}{!}{
\begin{tabular}{lcc}
\toprule
\text{Factor} & \text{Harmlessness} ($\uparrow$) & \text{Helpfulness} ($\uparrow$) \\
\hline
\multicolumn{3}{l}{\textit{Demonstration Type}} \\
\hline
\quad J         & 0.489& 0.501
\\
\quad J+IT      & 0.457& 0.492
\\
\quad J+IT+RR   & 0.469& 0.510\\
\quad J+RR      & \textbf{0.505}& \textbf{0.537}\\
\hline
\multicolumn{3}{l}{\textit{Judge Model}} \\
\hline
\quad Claude~3.5      & \textbf{0.502}& 0.506
\\
\quad Claude~3.7      & 0.464& \textbf{0.517}
\\
\quad DeepSeek-R1     & 0.482& 0.503
\\
\quad Nova Premier    & 0.471& 0.513\\
\hline
\multicolumn{3}{l}{\textit{Demonstration Count}} \\
\hline
\quad 1-shot    & 0.444& 0.491
\\
\quad 2-shot    & 0.472& 0.511
\\
\quad 4-shot    & 0.500& \textbf{0.519}
\\
\quad 8-shot    & \textbf{0.503}& \textbf{0.519}\\
\bottomrule
\end{tabular}
}
\caption{Average simulation accuracy by demonstration type, model, and demonstration
count. Each block isolates one factor while averaging over the other two. Bold marks
the highest value per task within each block.}
\label{tab:demo_factors}
\end{table}

\paragraph{Evaluator-specific signal effects}
\cref{tab:demo_factors} shows consistent demonstration-type patterns across both datasets: retrospective reasoning is the most useful complementary signal, and interface telemetry the least.
\textbf{J+RR} achieves the highest average accuracy on both tasks ($0.505$ Harmlessness, $0.537$ Helpfulness), whereas \textbf{J+IT} is worst and underperforms judgments alone ($0.457$, $0.492$); even combined with reasoning, telemetry pulls \textbf{J+IT+RR} below J+RR on both tasks.
The demonstration-type effect is statistically robust (Friedman $\chi^2(3)=22.4$/$20.0$, $p<0.001$; J+RR\,$>$\,J+IT under Wilcoxon, $p\le0.012$, $d\ge0.85$; \cref{app:significance}).
\paragraph{Model and demonstration-count effects}
Both factors matter less than demonstration type. 
Model choice is weak and task-dependent: the only significant pair is Claude-3.5-Sonnet\,$>$\,Claude-3.7-Sonnet on Harmlessness ($p=0.002$), with no significant model differences on Helpfulness. 
Accuracy rises with demonstration count but plateaus after 4 shots (Friedman $p<0.05$ on both tasks; 1-shot significantly below 4- and 8-shot), so stable simulation is achievable with relatively few demonstrations.

\paragraph{Best overall configuration}
\cref{tab:table_best_config} highlights the strongest condition combinations across tasks. 
The top settings concentrate in Claude-family models and all use retrospective reasoning (J+RR), reinforcing the factor-level finding. 
We recommend a single configuration, \textbf{Claude-3.5-Sonnet with 8-shot J+RR}, which reaches $0.581$ on Harmlessness ($+9.9$ points over its Base Judge) and $0.558$ on Helpfulness ($+5.8$ points). This configuration significantly improves over its Base Judge on \emph{both} tasks (Wilcoxon $p=0.046$ and $p=0.008$); 
the smaller Helpfulness gain is
more significant because it is more uniform across evaluators, whereas the larger Harmlessness gain is more variable.
Claude-3.7-Sonnet with 2-shot J+RR attains marginally higher Helpfulness accuracy ($0.568$), but its gain over the Base Judge is not significant ($p=0.057$).



\begin{table}[t]
\centering
\resizebox{0.9\columnwidth}{!}{
\begin{tabular}{lcc}
\toprule
\text{Factor Comb.} & \text{Harmless} ($\uparrow$) & \text{Helpful} ($\uparrow$)\\
\midrule
\multicolumn{3}{l}{\textit{Claude-3.5-Sonnet}}\\
\hline
Base Judge          & $0.482_{0.124}$ & $0.500_{0.104}$    \\
J+RR (2-Shot)       & $0.532_{0.103}$ & $0.544_{0.091}$                           \\
J+RR (4-Shot)       & $0.571_{0.130}$ & $0.548_{0.081}$                           \\
J+RR (8-Shot)       & $\textbf{0.581}_{0.116}$ & $\textbf{0.558}_{0.068}$             \\
\midrule
\multicolumn{3}{l}{\textit{Claude-3.7-Sonnet}}\\
\hline
Base Judge           & $0.480_{0.140}$ & $0.506_{0.135}$     \\
J+RR (2-Shot)        & $0.460_{0.119}$ & $\underline{0.568}_{0.074}$                          \\
J+RR (4-Shot)        & $0.494_{0.148}$ & $0.557_{0.082}$                          \\
J+RR (8-Shot)        & $0.537_{0.131}$ & $0.558_{0.086}$                          \\


\hline
\end{tabular}
}
\caption{Top configuration performance in simulation accuracy (ACC) with standard-deviation subscript. We report the highest-performing combinations of model, demonstration type, and demonstration count on the Harmlessness and Helpfulness tasks. \textbf{Bold} marks the recommended configuration (Claude-3.5-Sonnet, 8-shot J+RR), which significantly improves over its Base Judge on both tasks ($p=0.046$ Harmless, $p=0.008$ Helpful). \underline{Underline} marks the numerically highest Helpfulness value (Claude-3.7-Sonnet, 2-shot J+RR), whose gain over the Base Judge is not significant ($p=0.057$).}
\label{tab:table_best_config}
\end{table}

\subsubsection{RQ3: Evaluator-Level Simulatability}
\label{subsec:teachability_patterns}
To understand why some evaluators are easier to simulate than others, we analyze per-evaluator accuracy and its relationship to task performance and judgment behavior.

\paragraph{Variation in per-evaluator fidelity}
Individual simulation accuracy varies widely. Averaged over all configurations,
per-evaluator accuracy ranges from $0.375$ to $0.565$ on Harmlessness and from $0.386$ to $0.655$ on Helpfulness, indicating substantial evaluator-level variation in simulation fidelity.

\paragraph{Genuine but modest individual capture}
Does this variation reflect genuine individual simulation, or just a better group
answer? The decisive test is \emph{deviation} items---where an annotator disagrees
with the item's consensus label ($20.2\%$ / $23.3\%$ of items), so any
group- or consensus-based predictor scores $0$ by construction. On these, \ours~still
recovers the annotator's exact label $0.367$ / $0.360$ of the time---accuracy
unattainable from group-level information---and its deviations are individual-aligned
in both incidence and direction (Appendix~\ref{app:error_analysis}). The captured
signal is nonetheless modest: \ours~exceeds a global majority-class baseline but
\emph{not} a per-evaluator one (each individual's own most frequent label;
$\Delta={-}0.019$, $p=0.95$ Harmlessness; $\Delta={+}0.042$, $p=0.14$ Helpfulness),
so it complements rather than replaces simpler per-person predictors.

\paragraph{Judgment style predicts difficulty}
Two facets of an evaluator's judgment style explain who is hard to simulate. 
First, \emph{neutral usage}: \texttt{Neutral} is modal in Harmlessness ($43.9\%$) but rarer in Helpfulness ($31.1\%$, vs.\ $34.0\%$ Prefer A / $34.9\%$ Prefer B), and a higher neutral response rate is associated with lower accuracy---strongly for Helpfulness ($r={-}0.894$) but only weakly for Harmlessness ($r={-}0.269$). 
Second, \emph{divergence from consensus}: an evaluator's deviation rate---the share of items where their label differs from the majority of the \emph{other} annotators\footnote{Leave-one-out, so an evaluator never contributes to the consensus they are scored against; an all-rater consensus yields the same negative direction but weaker estimates.}---is a significant negative predictor of accuracy on both tasks ($r={-}0.463$, $p=0.034$ Harmlessness; $r={-}0.676$, $p<0.001$ Helpfulness), and the negative association persists after partialling out neutral usage (Harmlessness $r={-}0.576$, $p<0.01$; Helpfulness $r={-}0.427$, $p=0.06$). 
Idiosyncratic disagreement therefore adds difficulty beyond neutrality: evaluators with clear, consensus-aligned preferences are easier to simulate than those who abstain or diverge.



\begin{figure}
    \centering
    \includegraphics[width=0.82\linewidth]{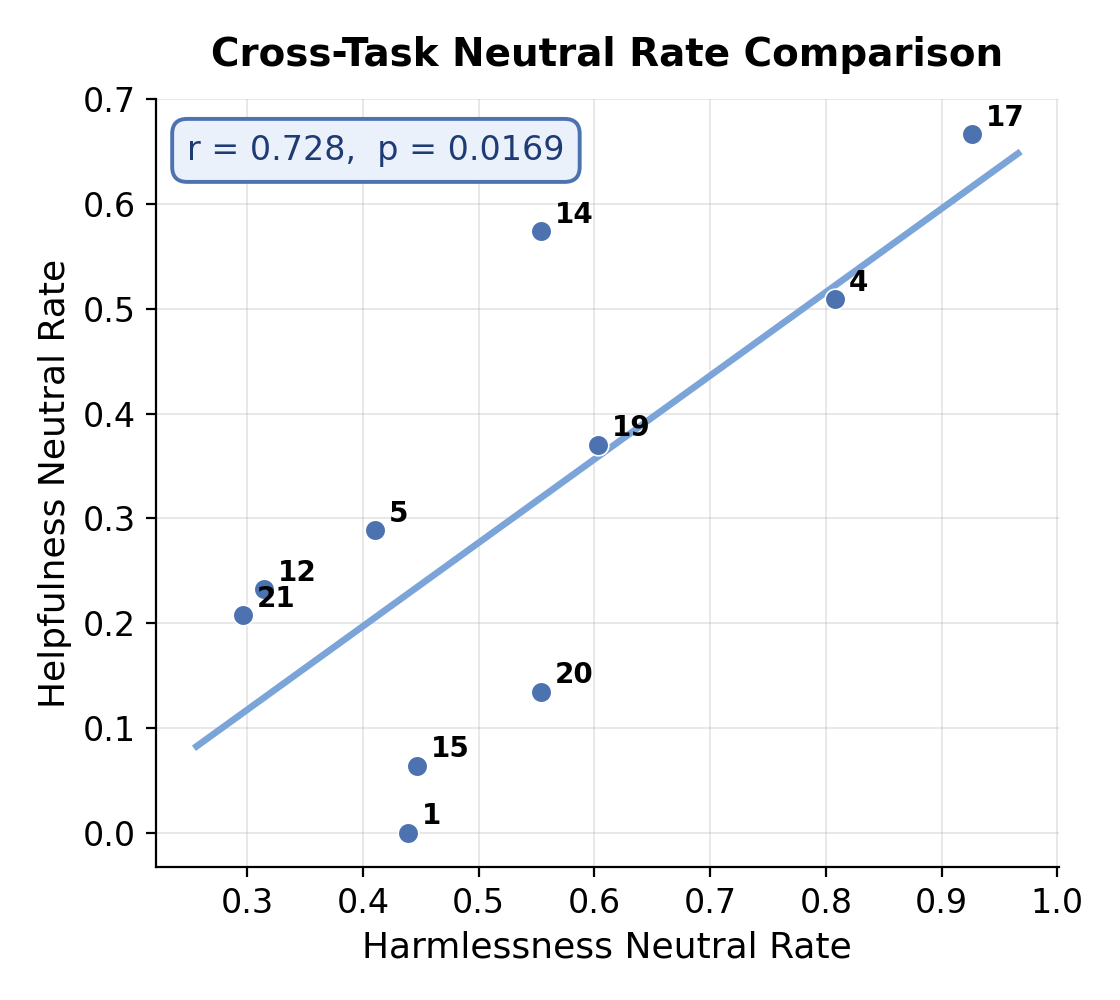}
    \caption{Evaluators' \texttt{Neutral}-selection rates are strongly correlated across tasks ($r=0.728$, $p<0.05$), whereas simulation fidelity itself is not ($r=0.181$). Each point is one of the 10 evaluators participating in both tasks; Neutral rates are computed on each evaluator's 60 validation items with a human majority label, so they can differ slightly from the full-data rates in \cref{fig:judgment-distribution}.
    }
    \label{fig:cross_task_simulability}
\end{figure}

\paragraph{A stable trait with a task-dependent effect}
For the 10 evaluators who completed both tasks, simulation fidelity itself does \emph{not} transfer across tasks ($r=0.181$, $p=0.616$). What transfers is the evaluator's \emph{neutral-usage tendency}: their \texttt{Neutral}-selection rate is strongly correlated across the two tasks ($r=0.728$, $p<0.05$; \cref{fig:cross_task_simulability}). 
Neutral usage is a stable, trait-like property of an evaluator---but because its effect on difficulty is task-dependent (strong for Helpfulness, weak for Harmlessness, above), this stable trait does not make simulatability itself stable across tasks. We interpret this cautiously given the limited overlap set ($n=10$).

\subsection{Discussion}
\label{sec:discussion}

\paragraph{Why does retrospective reasoning help more than interface telemetry?}
\label{subsec:why_reasoning_helps}
Retrospective reasoning likely helps because it externalizes the internal logic of a decision---how evaluators interpret context, weigh trade-offs, and justify uncertainty---in a text form LLMs can readily use, aligning the model with not only \emph{what} was judged but \emph{why}.
Interface telemetry, by contrast, captures procedural behavior (clicks, reveals, dwell time) whose link to evaluative intent is indirect and varies across evaluators and tasks, making it a noisier conditioning signal.
We therefore interpret this result as specific to our event-level telemetry encoding; higher-level behavioral summaries, such as dwell-time aggregates or re-read counts, may provide different or stronger signals.


\paragraph{Uneven but systematic simulatability}
\label{subsec:stable_simulability}

Because simulation difficulty tracks evaluator judgment style rather than noise (\cref{subsec:teachability_patterns}), it is unlikely to close with more of the same signals: current methods reproduce decisive, consensus-aligned styles more readily than uncertain or idiosyncratic ones, and the individual signal they capture, though genuine, does not yet exceed a person's own label tendency.
Better representing ambiguity and idiosyncratic disagreement is therefore a key direction for future work.

\paragraph{Implications for evaluator-specific simulation}
\label{subsec:implications}

These findings suggest that evaluator-specific simulation can complement consensus toward modeling meaningful individual variation. 
Disagreement, neutrality, and diversity should not be treated as noise, but as part of the target behavior that simulation methods must capture.

Practically, complementary signals show a clear cost--benefit asymmetry: retrospective reasoning is more costly to collect than interface telemetry---roughly five times more time per item---but contributes substantially more to simulation fidelity and interpretability (per-stage annotation times in \cref{app:cost}).
Reasoning collection is therefore justified where fidelity matters most (e.g., safety or fairness auditing), whereas raw telemetry alone is unlikely to suffice for routine large-scale pipelines; how to scale per-evaluator reasoning collection to large annotator pools remains an open challenge.



\section{Conclusion and Future Work}
\label{sec:conclusion}

We introduce \ours, an evaluator-specific simulation framework. \ours models evaluator-specific judgments by combining categorical preferences, retrospective reasoning, and interface telemetry through in-context learning.
Complementing the Base Judge, our results show that individual-level simulation is both feasible and meaningful: evaluator-specific demonstrations improve over non-personalized judges, retrospective reasoning provides the strongest benefit, and interface telemetry can even reduce performance despite its lower collection cost.
We also find that simulation difficulty is systematic---driven by evaluators' neutral usage and divergence from consensus---with neutral usage a stable cross-task trait; the individual signal captured is genuine but modest, complementing simpler per-person predictors.
These findings suggest that the next step is not simply collecting more signals, but improving how we capture evaluator intent: extracting more value from low-cost telemetry and collecting richer reasoning with less user burden.
More broadly, \ours~illustrates a shift in LLM-as-Judge from producing pooled labels to simulating how individuals evaluate, supporting systems that are both scalable and faithful to the inherent diversity in human judgment.
Beyond evaluation, such individually grounded simulations could provide fine-grained per-evaluator preference signals for reward modeling under heterogeneous human preferences~\cite{rame2023rewarded,pmlr-v235-chakraborty24b}, complementing synthetic-persona datasets constructed from aggregate demographic distributions~\cite{nemotron_personas_usa, riaz-etal-2025-metasynth, cisar2025pilotsteeringsyntheticdata, arannil-etal-2023-adeqa}.
Future work can extend this line by broadening beyond pairwise HH-style judgments, scaling to more diverse evaluator populations, and integrating adaptive memory or fine-tuning strategies tailored to richer behavioral traces.


\section*{Limitations} 
In this work, we focus on pairwise preference judgments over conversational follow-up responses in the Helpful and Harmless setting.
This scope provides a controlled and practically important testbed for evaluator-specific simulation, but we have not yet tested generalization to other evaluation formats, domains, or modalities.
Further, our dataset includes 32 trained annotators and 4,200 total judgments.
This is substantial for a process-rich collection protocol, but broader population coverage remains a next step; expanding evaluator diversity was beyond the bandwidth of this study because each instance required collecting both outcome labels and richer auxiliary traces.
\ours~uses in-context learning with sampled demonstrations as a strong and transparent baseline.
Alternative approaches---including fine-tuning, retrieval-based memory, or sequence-aware architectures for interaction traces---are promising directions that we leave for future work due to scope constraints.
Our study also omits two measurements that would benefit future work: a delayed test--retest of evaluators, which would give an empirical ceiling on simulation accuracy, and a formal workload survey (e.g., NASA-TLX) to quantify annotator fatigue.
Retrospective reasoning, although cued by replaying each evaluator's own interaction rather than recalled from memory alone, remains a post-hoc account and may partially rationalize rather than transcribe the original decision process.
Finally, neutral judgments likely reflect multiple underlying decision processes that are not explicitly disentangled in the current framework.
Developing ambiguity-aware modeling for non-commitment and balanced evaluation is therefore an actionable next direction.

\section*{Ethical Considerations}
Our prompt is released for reproducibility, but \ours~may be potentially misused when treated as a replacement for real human evaluators.
Because uncertainty and borderline cases remain challenging for \ours~to simulate reliably, overreliance on \ours~could misrepresent edge cases and bias downstream decisions. Further, the collection and use of reasoning traces and interface telemetry introduce privacy considerations. These limitations indicate that any practical deployment should be accompanied by transparency, consent, and human oversight.













\bibliography{custom}

@article{bai2022constitutional,
  title={Constitutional {AI}: Harmlessness from {AI} feedback},
    author={Bai, Yuntao and Kadavath, Saurav and Kundu, Sandipan and Askell, Amanda and Kernion, Jackson and Jones, Andy and Chen, Anna and Goldie, Anna and Mirhoseini, Azalia and McKinnon, Cameron and Chen, Carol and Olsson, Catherine and Olah, Christopher and Hernandez, Danny and Drain, Dawn and Ganguli, Deep and Li, Dustin and Tran-Johnson, Eli and Perez, Ethan and Kerr, Jamie and Mueller, Jared and Ladish, Jeffrey and Landau, Joshua and Ndousse, Kamal and Lukosuite, Kamile and Lovitt, Liane and Sellitto, Michael and Elhage, Nelson and Schiefer, Nicholas and Mercado, Noemi and DasSarma, Nova and Lasenby, Robert and Larson, Robin and Ringer, Sam and Johnston, Scott and Kravec, Shauna and {El Showk}, Sheer and Fort, Stanislav and Lanham, Tamera and Telleen-Lawton, Timothy and Conerly, Tom and Henighan, Tom and Hume, Tristan and Bowman, Samuel R. and Hatfield-Dodds, Zac and Mann, Ben and Amodei, Dario and Joseph, Nicholas and McCandlish, Sam and Brown, Tom and Kaplan, Jared},
  journal={arXiv preprint arXiv:2212.08073},
  year={2022}
}

@article{zheng2023judging,
  title={Judging {LLM}-as-a-judge with {MT-Bench} and {Chatbot Arena}},
  author={Zheng, Lianmin and Chiang, Wei-Lin and Sheng, Ying and Zhuang, Siyuan and Wu, Zhanghao and Zhuang, Yonghao and Lin, Zi and Li, Zhuohan and Li, Dacheng and Xing, Eric and Zhang, Hao and Gonzalez, Joseph and Stoica, Ion},
  journal={Advances in neural information processing systems},
  volume={36},
  pages={46595--46623},
  year={2023}
}

@inproceedings{snow2008cheap,
  title={Cheap and fast--but is it good? evaluating non-expert annotations for natural language tasks},
  author={Snow, Rion and O'Connor, Brendan and Jurafsky, Dan and Ng, Andrew Y},
  booktitle={Proceedings of the 2008 conference on empirical methods in natural language processing},
  pages={254--263},
  year={2008}
}

@inproceedings{lee2023rlaif,
author = {Lee, Harrison and Phatale, Samrat and Mansoor, Hassan and Mesnard, Thomas and Ferret, Johan and Lu, Kellie and Bishop, Colton and Hall, Ethan and Carbune, Victor and Rastogi, Abhinav and Prakash, Sushant},
title = {{RLAIF} vs. {RLHF}: scaling reinforcement learning from human feedback with AI feedback},
year = {2024},
publisher = {JMLR.org},
booktitle = {Proceedings of the 41st International Conference on Machine Learning},
articleno = {1071},
numpages = {28},
location = {Vienna, Austria},
series = {ICML'24}
}

@inproceedings{williams2017deja,
  title={Deja vu: Characterizing worker reliability using task consistency},
  author={Williams, Alex and Goh, Joslin and Willis, Charlie and Ellison, Aaron and Brusuelas, James and Davis, Charles and Law, Edith},
  booktitle={Proceedings of the AAAI Conference on Human Computation and Crowdsourcing},
  volume={5},
  pages={197--205},
  year={2017}
}

@inproceedings{liu-etal-2023-g,
    title = "{G}-Eval: {NLG} Evaluation using Gpt-4 with Better Human Alignment",
    author = "Liu, Yang  and
      Iter, Dan  and
      Xu, Yichong  and
      Wang, Shuohang  and
      Xu, Ruochen  and
      Zhu, Chenguang",
    editor = "Bouamor, Houda  and
      Pino, Juan  and
      Bali, Kalika",
    booktitle = "Proceedings of the 2023 Conference on Empirical Methods in Natural Language Processing",
    month = dec,
    year = "2023",
    address = "Singapore",
    publisher = "Association for Computational Linguistics",
    url = "https://aclanthology.org/2023.emnlp-main.153/",
    doi = "10.18653/v1/2023.emnlp-main.153",
    pages = "2511--2522"
}

@article{stiennon2020learning,
  title={Learning to summarize with human feedback},
  author={Stiennon, Nisan and Ouyang, Long and Wu, Jeffrey and Ziegler, Daniel and Lowe, Ryan and Voss, Chelsea and Radford, Alec and Amodei, Dario and Christiano, Paul F},
  journal={Advances in neural information processing systems},
  volume={33},
  pages={3008--3021},
  year={2020}
}

@article{nakano2021webgpt,
  title={{WebGPT}: Browser-assisted question-answering with human feedback},
  author={Reiichiro Nakano and Jacob Hilton and Suchir Balaji and Jeff Wu and Ouyang Long and Christina Kim and Christopher Hesse and Shantanu Jain and Vineet Kosaraju and William Saunders and Xu Jiang and Karl Cobbe and Tyna Eloundou and Gretchen Krueger and Kevin Button and Matthew Knight and Benjamin Chess and John Schulman},
  journal={ArXiv},
  year={2021},
  volume={abs/2112.09332},
  url={https://api.semanticscholar.org/CorpusID:245329531}
}

@InProceedings{pmlr-v162-ethayarajh22a,
  title = 	 {Understanding Dataset Difficulty with $\mathcal{V}$-Usable Information},
  author =       {Ethayarajh, Kawin and Choi, Yejin and Swayamdipta, Swabha},
  booktitle = 	 {Proceedings of the 39th International Conference on Machine Learning},
  pages = 	 {5988--6008},
  year = 	 {2022},
  editor = 	 {Chaudhuri, Kamalika and Jegelka, Stefanie and Song, Le and Szepesvari, Csaba and Niu, Gang and Sabato, Sivan},
  volume = 	 {162},
  series = 	 {Proceedings of Machine Learning Research},
  month = 	 {17--23 Jul},
  publisher = {PMLR},
}

@inproceedings{wang2023large,
    title = "Large Language Models are not Fair Evaluators",
    author = "Wang, Peiyi  and
      Li, Lei  and
      Chen, Liang  and
      Cai, Zefan  and
      Zhu, Dawei  and
      Lin, Binghuai  and
      Cao, Yunbo  and
      Kong, Lingpeng  and
      Liu, Qi  and
      Liu, Tianyu  and
      Sui, Zhifang",
    editor = "Ku, Lun-Wei  and
      Martins, Andre  and
      Srikumar, Vivek",
    booktitle = "Proceedings of the 62nd Annual Meeting of the Association for Computational Linguistics (Volume 1: Long Papers)",
    month = aug,
    year = "2024",
    address = "Bangkok, Thailand",
    publisher = "Association for Computational Linguistics",
    url = "https://aclanthology.org/2024.acl-long.511/",
    doi = "10.18653/v1/2024.acl-long.511",
    pages = "9440--9450"
}

@inproceedings{
ye2024justiceprejudicequantifyingbiases,
title={Justice or Prejudice? Quantifying Biases in {LLM}-as-a-Judge},
author={Jiayi Ye and Yanbo Wang and Yue Huang and Dongping Chen and Qihui Zhang and Nuno Moniz and Tian Gao and Werner Geyer and Chao Huang and Pin-Yu Chen and Nitesh V Chawla and Xiangliang Zhang},
booktitle={NeurIPS Safe Generative AI Workshop 2024},
year={2024},
url={https://openreview.net/forum?id=wtscPS2zJH}
}

@inproceedings{shi2024judging,
  title={Judging the judges: A systematic study of position bias in {LLM}-as-a-judge},
  author={Shi, Lin and Ma, Chiyu and Liang, Wenhua and Diao, Xingjian and Ma, Weicheng and Vosoughi, Soroush},
  booktitle={Proceedings of the 14th International Joint Conference on Natural Language Processing and the 4th Conference of the Asia-Pacific Chapter of the Association for Computational Linguistics},
  pages={292--314},
  year={2025}
}

@article{bradley1952rank,
  title={Rank analysis of incomplete block designs: {I}. {T}he method of paired comparisons},
  author={Bradley, Ralph Allan and Terry, Milton E},
  journal={Biometrika},
  volume={39},
  number={3/4},
  pages={324--345},
  year={1952},
  publisher={JSTOR}
}

@inproceedings{shaikh2025creating,
  title={Creating general user models from computer use},
  author={Shaikh, Omar and Sapkota, Shardul and Rizvi, Shan and Horvitz, Eric and Park, Joon Sung and Yang, Diyi and Bernstein, Michael S},
  booktitle={Proceedings of the 38th Annual ACM Symposium on User Interface Software and Technology},
  pages={1--23},
  year={2025}
}

@misc{wang2025opera,
      title={{OPeRA}: A Dataset of Observation, Persona, Rationale, and Action for Evaluating LLMs on Human Online Shopping Behavior Simulation},
      author={Ziyi Wang and Yuxuan Lu and Wenbo Li and Amirali Amini and Bo Sun and Yakov Bart and Weimin Lyu and Jiri Gesi and Tian Wang and Jing Huang and Yu Su and Upol Ehsan and Malihe Alikhani and Toby Jia-Jun Li and Lydia Chilton and Dakuo Wang},
      year={2025},
      eprint={2506.05606},
      archivePrefix={arXiv},
      primaryClass={cs.CL},
      url={https://arxiv.org/abs/2506.05606},
}

@book{payne1993adaptive,
  title={The Adaptive Decision Maker},
  author={Payne, John W. and Bettman, James R. and Johnson, Eric J.},
  year={1993},
  publisher={Cambridge University Press},
  address={Cambridge}
}

@inproceedings{stumpf2007toward,
  title={Toward Harnessing User Feedback for Machine Learning},
  author={Stumpf, Simone and Rajaram, Vidya and Li, Lida and Burnett, Margaret and Dietterich, Thomas and Sullivan, Erin and Drummond, Russell and Herlocker, Jonathan},
  booktitle={Proceedings of the 12th International Conference on Intelligent User Interfaces},
  year={2007},
  pages={82--91}
}

@inproceedings{wiegreffe2021teach,
  title={Teach Me to Explain: A Review of Datasets for Explainable Natural Language Processing},
  author={Wiegreffe, Sarah and Marasovi{\'c}, Ana},
  booktitle={Proceedings of the Neural Information Processing Systems Track on Datasets and Benchmarks},
  year={2021}
}

@inproceedings{ouyang2022training,
  title={Training Language Models to Follow Instructions with Human Feedback},
  author={Ouyang, Long and Wu, Jeff and Jiang, Xu and Almeida, Diogo and Wainwright, Carroll L. and Mishkin, Pamela and Zhang, Chong and Agarwal, Sandhini and Slama, Katarina and Ray, Alex and Schulman, John and Hilton, Jacob and Kelton, Fraser and Miller, Luke and Simens, Maddie and Askell, Amanda and Welinder, Peter and Christiano, Paul F. and Leike, Jan and Lowe, Ryan},
  booktitle={Advances in Neural Information Processing Systems},
  year={2022},
  volume={35},
  publisher={Curran Associates, Inc.}
}

@article{bai2022training,
  title={Training a Helpful and Harmless Assistant with Reinforcement Learning from Human Feedback},
  author={Bai, Yuntao and Jones, Andy and Ndousse, Kamal and Askell, Amanda and Chen, Anna and DasSarma, Nova and Drain, Dawn and Fort, Stanislav and Ganguli, Deep and Henighan, Tom and Joseph, Nicholas and Kadavath, Saurav and Kernion, Jackson and Conerly, Tom and El-Showk, Sheer and Elhage, Nelson and Hatfield-Dodds, Zac and Hernandez, Danny and Hume, Tristan and Johnston, Scott and Kravec, Shauna and Lovitt, Liane and Nanda, Neel and Olsson, Catherine and Amodei, Dario and Brown, Tom and Clark, Jack and McCandlish, Sam and Olah, Chris and Mann, Ben and Kaplan, Jared},
  journal={arXiv preprint arXiv:2204.05862},
  year={2022}
}

@article{ford1989process,
  title={Process Tracing Methods: Contributions, Problems, and Neglected Research Questions},
  author={Ford, J. Kevin and Schmitt, Neal and Schechtman, Susan L. and Hults, Brian M. and Doherty, Mary L.},
  journal={Organizational Behavior and Human Decision Processes},
  year={1989},
  volume={43},
  number={1},
  pages={75--117}
}

@inproceedings{guo2010ready,
  title={Ready to Buy or Just Browsing? Detecting Web Searcher Goals from Interaction Data},
  author={Guo, Qi and Agichtein, Eugene},
  booktitle={Proceedings of the 33rd International ACM SIGIR Conference on Research and Development in Information Retrieval},
  year={2010},
  pages={130--137}
}

@inproceedings{liu2010understanding,
  title={Understanding Web Browsing Behaviors through Weibull Analysis of Dwell Time},
  author={Liu, Chao and White, Ryen W. and Dumais, Susan},
  booktitle={Proceedings of the 33rd International ACM SIGIR Conference on Research and Development in Information Retrieval},
  year={2010},
  pages={379--386}
}

@book{ericsson1993protocol,
  title={Protocol Analysis: Verbal Reports as Data},
  author={Ericsson, K. Anders and Simon, Herbert A.},
  year={1993},
  publisher={MIT Press},
  address={Cambridge, MA}
}

@article{vansomeren1994think,
  title={The think aloud method: a practical approach to modelling cognitive processes},
  author={Maarten van Someren and Yvonne Barnard and Jacobijn Sandberg},
  journal={Knowledge Based Systems},
  year={1994},
  url={https://api.semanticscholar.org/CorpusID:61817892}
}

@inproceedings{shankar2024validates,
  title={Who validates the validators? {A}ligning {LLM}-assisted evaluation of {LLM} outputs with human preferences},
  author={Shankar, Shreya and Zamfirescu-Pereira, JD and Hartmann, Bj{\"o}rn and Parameswaran, Aditya and Arawjo, Ian},
  booktitle={Proceedings of the 37th Annual ACM Symposium on User Interface Software and Technology},
  pages={1--14},
  year={2024}
}

@inproceedings{lambert2024rewardbench,
    title = "{R}eward{B}ench: Evaluating Reward Models for Language Modeling",
    author = "Lambert, Nathan  and
      Pyatkin, Valentina  and
      Morrison, Jacob  and
      Miranda, LJ  and
      Lin, Bill Yuchen  and
      Chandu, Khyathi  and
      Dziri, Nouha  and
      Kumar, Sachin  and
      Zick, Tom  and
      Choi, Yejin  and
      Smith, Noah A.  and
      Hajishirzi, Hannaneh",
    editor = "Chiruzzo, Luis  and
      Ritter, Alan  and
      Wang, Lu",
    booktitle = "Findings of the Association for Computational Linguistics: NAACL 2025",
    month = apr,
    year = "2025",
    address = "Albuquerque, New Mexico",
    publisher = "Association for Computational Linguistics",
    url = "https://aclanthology.org/2025.findings-naacl.96/",
    doi = "10.18653/v1/2025.findings-naacl.96",
    pages = "1755--1797",
    ISBN = "979-8-89176-195-7"
}

@inproceedings{wei2022chain,
author = {Wei, Jason and Wang, Xuezhi and Schuurmans, Dale and Bosma, Maarten and Ichter, Brian and Xia, Fei and Chi, Ed H. and Le, Quoc V. and Zhou, Denny},
title = {Chain-of-thought prompting elicits reasoning in large language models},
year = {2022},
isbn = {9781713871088},
publisher = {Curran Associates Inc.},
address = {Red Hook, NY, USA},
booktitle = {Proceedings of the 36th International Conference on Neural Information Processing Systems},
articleno = {1800},
numpages = {14},
location = {New Orleans, LA, USA},
series = {NIPS '22}
}

@article{kojima2022large,
  title={Large Language Models are Zero-Shot Reasoners},
  author={Kojima, Takeshi and Gu, Shixiang Shane and Reid, Machel and Matsuo, Yutaka and Iwasawa, Yusuke},
  journal={Advances in Neural Information Processing Systems},
  volume={35},
  pages={22199--22213},
  year={2022}
}

@inproceedings{dong2022survey,
    title = "A Survey on In-context Learning",
    author = "Dong, Qingxiu  and
      Li, Lei  and
      Dai, Damai  and
      Zheng, Ce  and
      Ma, Jingyuan  and
      Li, Rui  and
      Xia, Heming  and
      Xu, Jingjing  and
      Wu, Zhiyong  and
      Chang, Baobao  and
      Sun, Xu  and
      Li, Lei  and
      Sui, Zhifang",
    editor = "Al-Onaizan, Yaser  and
      Bansal, Mohit  and
      Chen, Yun-Nung",
    booktitle = "Proceedings of the 2024 Conference on Empirical Methods in Natural Language Processing",
    month = nov,
    year = "2024",
    address = "Miami, Florida, USA",
    publisher = "Association for Computational Linguistics",
    url = "https://aclanthology.org/2024.emnlp-main.64/",
    doi = "10.18653/v1/2024.emnlp-main.64",
    pages = "1107--1128"
}

@article{rame2023rewarded,
  title={Rewarded soups: towards {P}areto-optimal alignment by interpolating weights fine-tuned on diverse rewards},
  author={Rame, Alexandre and Couairon, Guillaume and Dancette, Corentin and Gaya, Jean-Baptiste and Shukor, Mustafa and Soulier, Laure and Cord, Matthieu},
  journal={Advances in Neural Information Processing Systems},
  volume={36},
  pages={71095--71134},
  year={2023}
}

@InProceedings{pmlr-v235-chakraborty24b,
  title = 	 {{M}ax{M}in-{RLHF}: Alignment with Diverse Human Preferences},
  author =       {Chakraborty, Souradip and Qiu, Jiahao and Yuan, Hui and Koppel, Alec and Manocha, Dinesh and Huang, Furong and Bedi, Amrit and Wang, Mengdi},
  booktitle = 	 {Proceedings of the 41st International Conference on Machine Learning},
  pages = 	 {6116--6135},
  year = 	 {2024},
  editor = 	 {Salakhutdinov, Ruslan and Kolter, Zico and Heller, Katherine and Weller, Adrian and Oliver, Nuria and Scarlett, Jonathan and Berkenkamp, Felix},
  volume = 	 {235},
  series = 	 {Proceedings of Machine Learning Research},
  month = 	 {21--27 Jul},
  publisher =    {PMLR},
  url = 	 {https://proceedings.mlr.press/v235/chakraborty24b.html}
}

@inproceedings{dong-etal-2024-llm,
    title = "Can {LLM} be a Personalized Judge?",
    author = "Dong, Yijiang River  and
      Hu, Tiancheng  and
      Collier, Nigel",
    editor = "Al-Onaizan, Yaser  and
      Bansal, Mohit  and
      Chen, Yun-Nung",
    booktitle = "Findings of the Association for Computational Linguistics: EMNLP 2024",
    month = nov,
    year = "2024",
    address = "Miami, Florida, USA",
    publisher = "Association for Computational Linguistics",
    url = "https://aclanthology.org/2024.findings-emnlp.592/",
    doi = "10.18653/v1/2024.findings-emnlp.592",
    pages = "10126--10141"
}

@article{chen2024chatgpt,
  title={How is {ChatGPT}’s behavior changing over time?},
  author={Chen, Lingjiao and Zaharia, Matei and Zou, James},
  journal={Harvard Data Science Review},
  volume={6},
  number={2},
  year={2024},
  publisher={The MIT Press}
}

@article{van2021human,
  title={Human evaluation of automatically generated text: Current trends and best practice guidelines},
  author={Van der Lee, Chris and Gatt, Albert and Van Miltenburg, Emiel and Krahmer, Emiel},
  journal={Computer Speech \& Language},
  volume={67},
  pages={101151},
  year={2021},
  publisher={Elsevier}
}

@article{davani2022dealing,
  title={Dealing with disagreements: Looking beyond the majority vote in subjective annotations},
  author={Davani, Aida Mostafazadeh and D{\'\i}az, Mark and Prabhakaran, Vinodkumar},
  journal={Transactions of the Association for Computational Linguistics},
  volume={10},
  pages={92--110},
  year={2022},
  publisher={MIT Press One Rogers Street, Cambridge, MA 02142-1209, USA journals-info~…}
}

@InProceedings{zhang2024diverging,
  title = 	 {Diverging Preferences: When do Annotators Disagree and do Models Know?},
  author =       {Zhang, Michael Jq and Wang, Zhilin and Hwang, Jena D. and Dong, Yi and Delalleau, Olivier and Choi, Yejin and Choi, Eunsol and Ren, Xiang and Pyatkin, Valentina},
  booktitle = 	 {Proceedings of the 42nd International Conference on Machine Learning},
  pages = 	 {76193--76212},
  year = 	 {2025},
  editor = 	 {Singh, Aarti and Fazel, Maryam and Hsu, Daniel and Lacoste-Julien, Simon and Berkenkamp, Felix and Maharaj, Tegan and Wagstaff, Kiri and Zhu, Jerry},
  volume = 	 {267},
  series = 	 {Proceedings of Machine Learning Research},
  month = 	 {13--19 Jul},
  publisher =    {PMLR},
  url = 	 {https://proceedings.mlr.press/v267/zhang25bx.html}
}

@inproceedings{fleisig2023majority,
  title={When the majority is wrong: Modeling annotator disagreement for subjective tasks},
  author={Fleisig, Eve and Abebe, Rediet and Klein, Dan},
  booktitle={Proceedings of the 2023 Conference on Empirical Methods in Natural Language Processing},
  pages={6715--6726},
  year={2023}
}

@inproceedings{basile-etal-2021-need,
    title = "We Need to Consider Disagreement in Evaluation",
    author = "Basile, Valerio  and
      Fell, Michael  and
      Fornaciari, Tommaso  and
      Hovy, Dirk  and
      Paun, Silviu  and
      Plank, Barbara  and
      Poesio, Massimo  and
      Uma, Alexandra",
    editor = "Church, Kenneth  and
      Liberman, Mark  and
      Kordoni, Valia",
    booktitle = "Proceedings of the 1st Workshop on Benchmarking: Past, Present and Future",
    month = aug,
    year = "2021",
    address = "Online",
    publisher = "Association for Computational Linguistics",
    url = "https://aclanthology.org/2021.bppf-1.3/",
    doi = "10.18653/v1/2021.bppf-1.3",
    pages = "15--21"
}

@article{frenda2025perspectivist,
  title={Perspectivist approaches to natural language processing: a survey},
  author={Frenda, Simona and Abercrombie, Gavin and Basile, Valerio and Pedrani, Alessandro and Panizzon, Raffaella and Cignarella, Alessandra Teresa and Marco, Cristina and Bernardi, Davide},
  journal={Language Resources and Evaluation},
  volume={59},
  number={2},
  pages={1719--1746},
  year={2025},
  publisher={Springer}
}

@inproceedings{jiang-etal-2025-language,
    title = "Can Language Models Reason about Individualistic Human Values and Preferences?",
    author = "Jiang, Liwei  and
      Sorensen, Taylor  and
      Levine, Sydney  and
      Choi, Yejin",
    editor = "Che, Wanxiang  and
      Nabende, Joyce  and
      Shutova, Ekaterina  and
      Pilehvar, Mohammad Taher",
    booktitle = "Proceedings of the 63rd Annual Meeting of the Association for Computational Linguistics (Volume 1: Long Papers)",
    month = jul,
    year = "2025",
    address = "Vienna, Austria",
    publisher = "Association for Computational Linguistics",
    url = "https://aclanthology.org/2025.acl-long.336/",
    doi = "10.18653/v1/2025.acl-long.336",
    pages = "6757--6794",
    ISBN = "979-8-89176-251-0"
}

@inproceedings{movva2024annotation,
  title={Annotation alignment: Comparing {LLM} and human annotations of conversational safety},
  author={Movva, Rajiv and Koh, Pang Wei and Pierson, Emma},
  booktitle={Proceedings of the 2024 Conference on Empirical Methods in Natural Language Processing},
  pages={9048--9062},
  year={2024}
}

@inproceedings{wang2024learning,
    title = "Learning Personalized Alignment for Evaluating Open-ended Text Generation",
    author = "Wang, Danqing  and
      Yang, Kevin  and
      Zhu, Hanlin  and
      Yang, Xiaomeng  and
      Cohen, Andrew  and
      Li, Lei  and
      Tian, Yuandong",
    editor = "Al-Onaizan, Yaser  and
      Bansal, Mohit  and
      Chen, Yun-Nung",
    booktitle = "Proceedings of the 2024 Conference on Empirical Methods in Natural Language Processing",
    month = nov,
    year = "2024",
    address = "Miami, Florida, USA",
    publisher = "Association for Computational Linguistics",
    url = "https://aclanthology.org/2024.emnlp-main.737/",
    doi = "10.18653/v1/2024.emnlp-main.737",
    pages = "13274--13292"
}

@article{franco2011applying,
  title={Applying the decision moving window to risky choice: Comparison of eye-tracking and mouse-tracing methods},
  author={Franco-Watkins, Ana M and Johnson, Joseph G},
  journal={Judgment and Decision making},
  volume={6},
  number={8},
  pages={740--749},
  year={2011},
  publisher={Cambridge University Press}
}

@article{jasper2002mousetrace,
  title={{MouseTrace}: A better mousetrap for catching decision processes},
  author={Jasper, JD and Shapiro, Jennifer},
  journal={Behavior Research Methods, Instruments, \& Computers},
  volume={34},
  number={3},
  pages={364--374},
  year={2002},
  publisher={Springer}
}

@article{schulte2011role,
  title={The role of process data in the development and testing of process models of judgment and decision making},
  author={Schulte-Mecklenbeck, Michael and K{\"u}hberger, Anton and Ranyard, Rob},
  journal={Judgment and Decision making},
  volume={6},
  number={8},
  pages={733--739},
  year={2011},
  publisher={Cambridge University Press}
}

@article{camburu2018snli,
  title={e-{SNLI}: Natural language inference with natural language explanations},
  author={Camburu, Oana-Maria and Rockt{\"a}schel, Tim and Lukasiewicz, Thomas and Blunsom, Phil},
  journal={Advances in Neural Information Processing Systems},
  volume={31},
  year={2018}
}

@article{zhou2020towards,
  title={Towards interpretable natural language understanding with explanations as latent variables},
  author={Zhou, Wangchunshu and Hu, Jinyi and Zhang, Hanlin and Liang, Xiaodan and Sun, Maosong and Xiong, Chenyan and Tang, Jian},
  journal={Advances in Neural Information Processing Systems},
  volume={33},
  pages={6803--6814},
  year={2020}
}

@article{park2026llm,
  title={{LLM} agents grounded in self-reports enable general-purpose simulation of individuals},
  author={Park, Joon Sung and Zou, Carolyn Q and Kamphorst, Jonne and Egan, Niles and Shaw, Aaron and Hill, Benjamin Mako and Cai, Carrie and Morris, Meredith Ringel and Liang, Percy and Willer, Robb and others},
  journal={arXiv preprint arXiv:2411.10109},
  year={2026}
}

@inproceedings{hu2025inews,
  title={{iNews}: A multimodal dataset for modeling personalized affective responses to news},
  author={Hu, Tiancheng and Collier, Nigel},
  booktitle={Proceedings of the 63rd Annual Meeting of the Association for Computational Linguistics (Volume 1: Long Papers)},
  pages={25000--25040},
  year={2025}
}

@article{de2025tracing,
  title={Tracing How Annotators Think: Augmenting Preference Judgments with Reading Processes},
  author={de Langis, Karin and Walker, William and Le, Khanh Chi and Kang, Dongyeop},
  journal={arXiv preprint arXiv:2511.21912},
  year={2025}
}

@article{gu2026survey,
  title={A survey on {LLM}-as-a-judge},
  author={Gu, Jiawei and Jiang, Xuhui and Shi, Zhichao and Tan, Hexiang and Zhai, Xuehao and Xu, Chengjin and Li, Wei and Shen, Yinghan and Ma, Shengjie and Liu, Honghao and others},
  journal={The Innovation},
  volume={7},
  number={6},
  year={2026},
  publisher={Elsevier}
}

@article{li2024llms,
  title={{LLMs}-as-judges: a comprehensive survey on {LLM}-based evaluation methods},
  author={Li, Haitao and Dong, Qian and Chen, Junjie and Su, Huixue and Zhou, Yujia and Ai, Qingyao and Ye, Ziyi and Liu, Yiqun},
  journal={arXiv preprint arXiv:2412.05579},
  year={2024}
}

@software{nemotron_personas_usa,
  author = {Meyer, Yev and Corneil, Dane},
  title = {{Nemotron-Personas-USA}: Synthetic Personas Aligned to Real-World Distributions
},
  month = {June},
  year = {2025},
  url = {https://huggingface.co/datasets/nvidia/Nemotron-Personas-USA}
}

@inproceedings{riaz-etal-2025-metasynth,
    title = "{M}eta{S}ynth: Meta-Prompting-Driven Agentic Scaffolds for Diverse Synthetic Data Generation",
    author = "Riaz, Haris  and
      Bhabesh, Sourav Sanjukta  and
      Arannil, Vinayak  and
      Ballesteros, Miguel  and
      Horwood, Graham",
    editor = "Che, Wanxiang  and
      Nabende, Joyce  and
      Shutova, Ekaterina  and
      Pilehvar, Mohammad Taher",
    booktitle = "Findings of the Association for Computational Linguistics: ACL 2025",
    month = jul,
    year = "2025",
    address = "Vienna, Austria",
    publisher = "Association for Computational Linguistics",
    url = "https://aclanthology.org/2025.findings-acl.962/",
    doi = "10.18653/v1/2025.findings-acl.962",
    pages = "18770--18803",
    ISBN = "979-8-89176-256-5"
}

@misc{cisar2025pilotsteeringsyntheticdata,
      title={{PILOT}: Steering Synthetic Data Generation with Psychological \& Linguistic Output Targeting}, 
      author={Caitlin Cisar and Emily Sheffield and Joshua Drake and Alden Harrell and Subramanian Chidambaram and Nikita Nangia and Vinayak Arannil and Alex Williams},
      year={2025},
      eprint={2509.15447},
      archivePrefix={arXiv},
      primaryClass={cs.CL},
      url={https://arxiv.org/abs/2509.15447}, 
}

@inproceedings{arannil-etal-2023-adeqa,
    title = "{ADEQA}: A Question Answer based approach for joint {ADE}-Suspect Extraction using Sequence-To-Sequence Transformers",
    author = "Arannil, Vinayak  and
      Deb, Tomal  and
      Roy, Atanu",
    editor = "Demner-fushman, Dina  and
      Ananiadou, Sophia  and
      Cohen, Kevin",
    booktitle = "Proceedings of the 22nd Workshop on Biomedical Natural Language Processing and BioNLP Shared Tasks",
    month = jul,
    year = "2023",
    address = "Toronto, Canada",
    publisher = "Association for Computational Linguistics",
    url = "https://aclanthology.org/2023.bionlp-1.17/",
    doi = "10.18653/v1/2023.bionlp-1.17",
    pages = "206--214"
}

@inproceedings{he2025prompting,
author = {He, Zeyu and Naphade, Saniya and Huang, Ting-Hao Kenneth},
title = {Prompting in the Dark: Assessing Human Performance in Prompt Engineering for Data Labeling When Gold Labels Are Absent},
year = {2025},
isbn = {9798400713941},
publisher = {Association for Computing Machinery},
address = {New York, NY, USA},
url = {https://doi.org/10.1145/3706598.3714319},
doi = {10.1145/3706598.3714319},
booktitle = {Proceedings of the 2025 CHI Conference on Human Factors in Computing Systems},
articleno = {1195},
numpages = {33},
location = {
},
series = {CHI '25}
}

\appendix

\section{Data Collection}\label{app:data-collection}
\subsection{User Interface Implementation Details}\label{sec:implementation}
We implement \ours data collection system as a managed AWS service. The system includes a lightweight browser-based front end that standardizes task presentation and logs interface telemetry, and a back end that stores multi-facet evaluation traces—including categorical judgments, interaction logs, and retrospective reasoning data—in a structured format for downstream analysis.

\begin{figure*}
    \centering
    \includegraphics[width=0.85\linewidth]{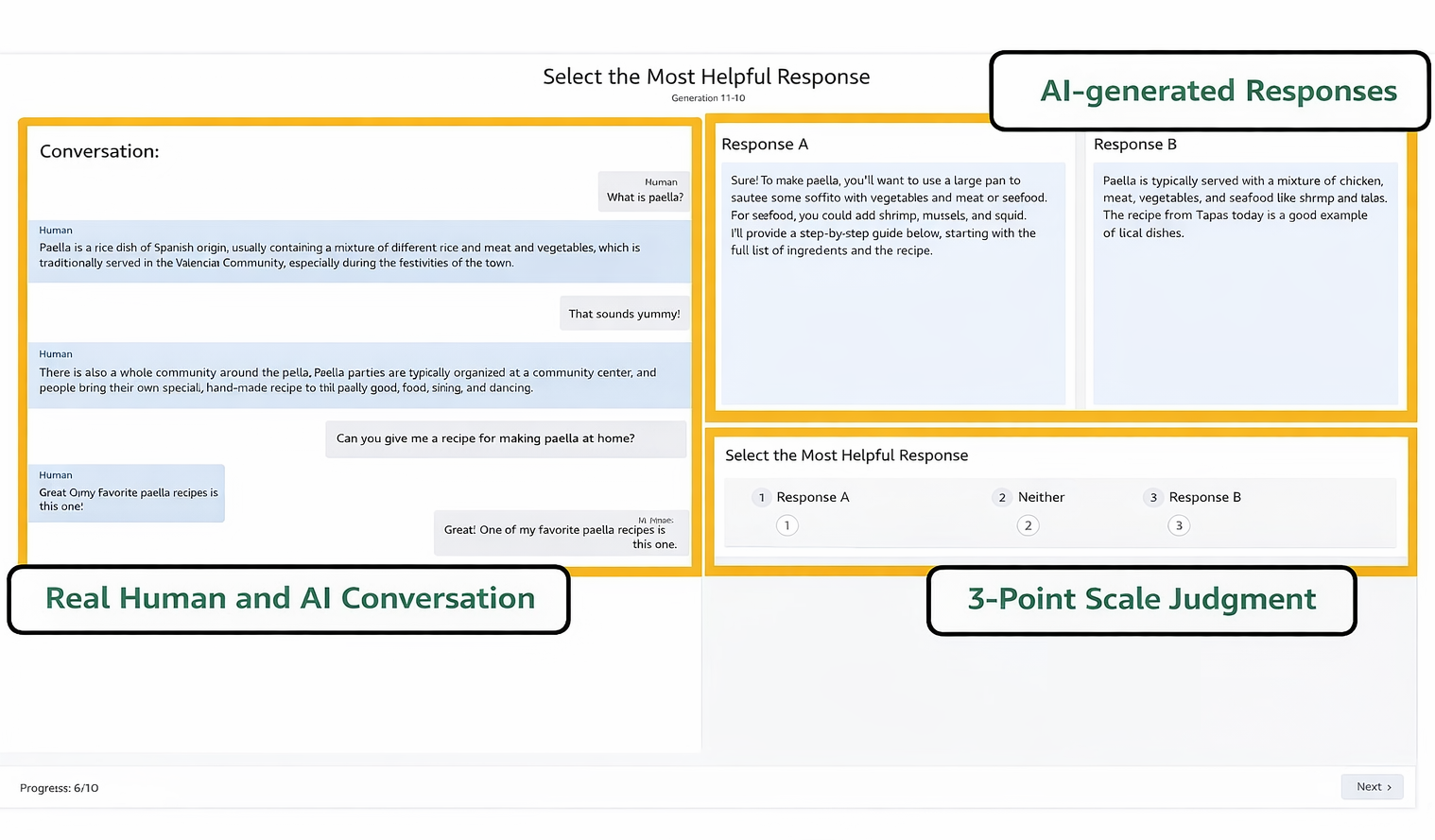}
    \caption{Stage 1 interface for preference judgment collection, showing the conversation context, two AI responses, and a three-point judgment scale.}
    \label{fig:ui-1}
\end{figure*}

\paragraph{Stage 1 Interface (\cref{fig:ui-1})}
The Stage 1 interface is designed for preference evaluation. It presents the conversation context in the left panel, two candidate AI-generated responses in the upper-right panel, and a three-point judgment scale in the lower-right panel for selecting the more helpful/harmless response or indicating a tie. Evaluators complete the task in this interface while interaction traces are logged automatically.

\begin{figure*}
    \centering
    \includegraphics[width=0.85\linewidth]{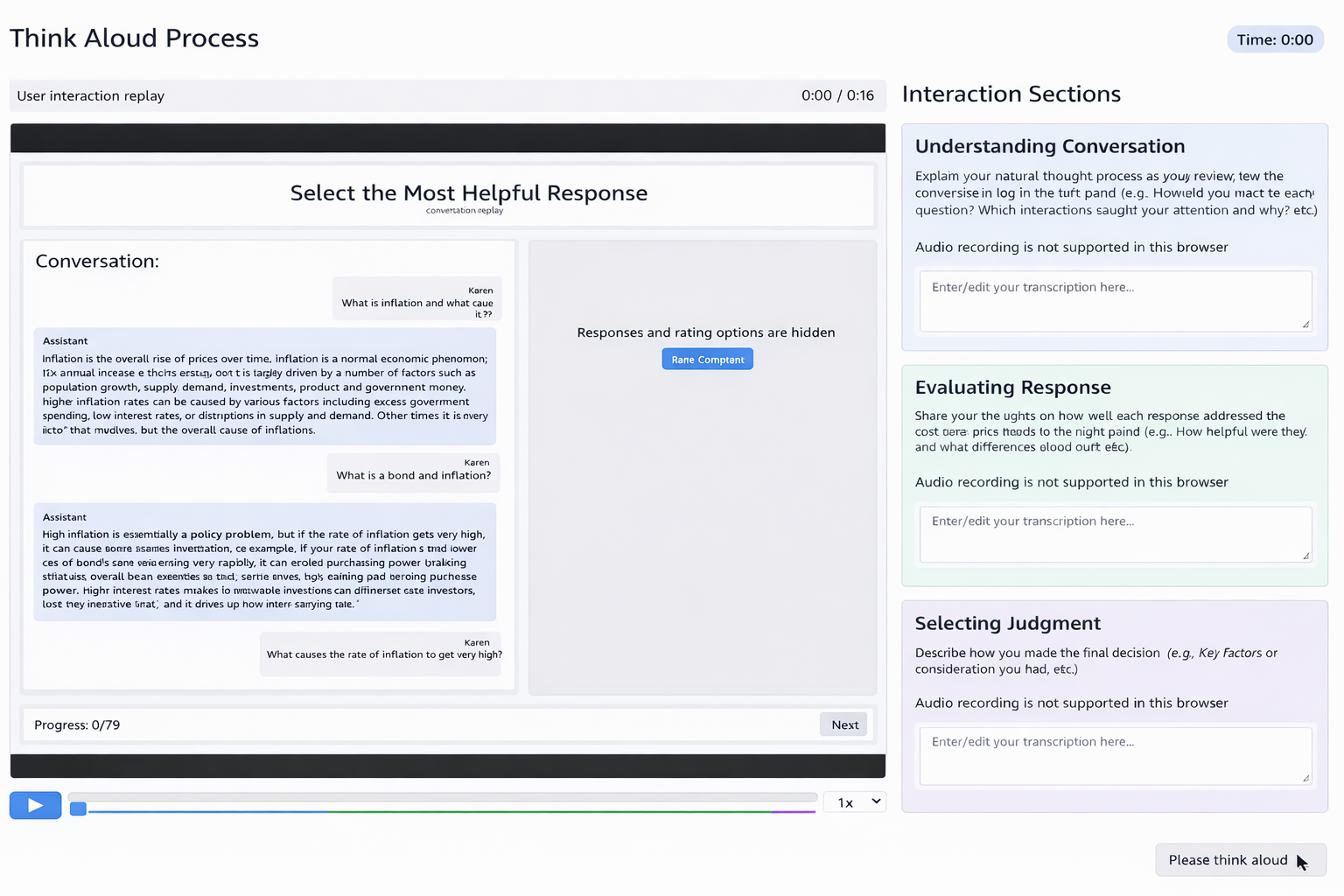}
    \caption{Stage 2 interface for retrospective reasoning, showing an interaction replay and structured think-aloud input fields.}
    \label{fig:ui-2}
\end{figure*}

\paragraph{Stage 2 Interface (\cref{fig:ui-2})}
The Stage 2 interface is designed for retrospective think-aloud collection. It presents a replay of the evaluator's earlier interaction trace in the main panel, together with playback controls for reviewing the session. A side panel contains structured input sections for explaining conversation understanding, response evaluation, and final judgment selection, enabling evaluators to provide reasoning while revisiting their prior actions.

\subsection{Dataset Selection and Assignment}
For our study, we randomly selected 700 unique (i.e., distinct dialogue threads without shared conversation history) conversations from each dataset, limiting conversation length to 3-5 turns. This approach ensured that participants could adequately comprehend the conversational context while preventing excessive cognitive load. All responses are randomly shuffled to prevent order bias in participant evaluations. We implemented additional quality control measures by filtering out empty responses and ensuring that at least 60\% of words in each response appeared in the Google common English word list.\footnote{\url{https://github.com/first20hours/google-10000-english/tree/master}}

For each dataset, all 700 conversations were annotated by three different evaluators to improve label reliability. 
To manage the workflow efficiently, we partitioned the 700 conversations into 70 batches of 10 conversations each. 
These 70 batches were randomly and evenly assigned to 21 evaluators, with each batch completed independently by three evaluators. To reduce potential order effects, the presentation order of conversations within each batch was randomly shuffled before evaluators began the task.

\subsection{User Study Procedure}
The user study procedure consisted of onboarding and tutorial training, followed by a two-stage annotation workflow.
Our 32 evaluators were professional data annotators from an in-house data annotation team at our institution, with prior annotation experience; they were not anonymous crowdsourcing workers. 
Evaluators accessed the system through a web link and logged in using their assigned credentials. All judgments were completed remotely on the evaluators' own laptops.

\paragraph{Onboarding} Evaluators were first introduced to the study goals, task procedures, and data handling practices. Informed consent was obtained prior to participation. Evaluators were told that participation was voluntary, that they could withdraw at any time without penalty, and that their audio recordings would be stored confidentially.

\paragraph{Tutorial and training} Evaluators were then introduced to the system workflow through either a live Zoom onboarding session or a recorded tutorial video, depending on when they joined the study. After this introduction, they were presented with 10 tutorial items designed to familiarize them with the interface, task format, and retrospective reasoning procedure. 
These tutorial items were reviewed to ensure that evaluators understood the workflow before beginning the main annotation tasks.

\paragraph{Main annotation tasks} 
Each evaluator completed 100 assigned evaluation instances, organized into 10 batches. 
Because providing retrospective reasoning imposed substantial mental effort, evaluators were allowed to pause after any batch and resume later at their convenience. 
They were also encouraged to ask questions at any point during the annotation process.




\section{Annotation Time}\label{app:cost}

We quantify the per-item annotation effort of each collection stage from the
interaction logs ($n=2{,}100$ tasks per dataset; 21 evaluators $\times$ 100 items).
\textbf{Stage 1} time (categorical judgment with interface telemetry, collected
together) is measured from the interface telemetry as hands-on time---from the start
of an item to its final \texttt{rating\_selected} event---with idle pauses longer than
$30$\,s removed so the figure reflects active engagement rather than breaks.
\textbf{Stage 2} time (retrospective think-aloud) is the duration of each evaluator's
reasoning sections (conversation understanding, response evaluation, and judgment).

\begin{table}[t]
\centering
\small
\resizebox{0.9\linewidth}{!}{\begin{tabular}{lcc}
\toprule
Stage & Harmlessness & Helpfulness \\
\midrule
Stage 1: judgment $+$ telemetry & $56.2$\,s & $57.4$\,s \\
\quad (median) & $48.0$\,s & $47.0$\,s \\
Stage 2: retrospective reasoning & $308.5$\,s & $275.3$\,s \\
\quad (median) & $280.6$\,s & $209.8$\,s \\
\midrule
Total per item (mean) & $364.7$\,s & $332.7$\,s \\
\bottomrule
\end{tabular}}
\caption{Mean (and median) per-item annotation time by stage, in seconds, over the
$2{,}100$ tasks per dataset. Stage-1 time is idle-adjusted active engagement; Stage-2
time is the think-aloud recording duration.}
\label{tab:app-cost}
\end{table}

\cref{tab:app-cost} reports the figures. Collecting a categorical judgment takes
under a minute per item ($\sim$56--57\,s), and \textbf{interface telemetry adds no
marginal cost} because it is logged passively during Stage~1. In contrast,
retrospective reasoning is roughly \textbf{five times} more expensive to collect
($\sim$4.6--5.1\,min per item in Stage~2), bringing the total to about $6$\,min per
item. This cost asymmetry underlies the practical trade-off discussed in
\S\ref{subsec:implications}: the most informative signal (retrospective reasoning) is
also by far the most expensive to collect, whereas the cheapest signal (telemetry)
contributes least to simulation fidelity.

\section{Demonstration Format}\label{app:demonstration-format}

\paragraph{Processed Interface Telemetry}\label{app:telemtry-format}
\Cref{tab:telemetry} presents the structured interface telemetry schema used in \ours demonstrations. 
\begin{table*}[h]
\centering
\caption{Structured interface telemetry schema used in PersonaJudge demonstrations. Each interaction trace is serialized as a sequence of typed events with event-specific fields.}
\label{tab:telemetry}
\scriptsize
\setlength{\tabcolsep}{4pt}
\begin{tabular}{@{}llllllll@{}}
\toprule
\textbf{Event Type} & \textbf{Field 1} & \textbf{Field 2} & \textbf{Field 3} & \textbf{Field 4} & \textbf{Field 5} & \textbf{Field 6} & \textbf{Field 7} \\
\midrule
\texttt{mouse\_move}      & \texttt{coord\_from}      & \texttt{coord\_to}       & \texttt{timestamp}      & \texttt{ui\_section}    & ---                      & ---                    & --- \\
\texttt{left\_click}      & \texttt{coordinate}       & \texttt{timestamp}       & \texttt{ui\_section}    & \texttt{element\_type}  & \texttt{text}            & ---                    & --- \\
\texttt{double\_click}    & \texttt{coordinate}       & \texttt{timestamp}       & \texttt{ui\_section}    & \texttt{element\_type}  & \texttt{selected\_text}  & ---                    & --- \\
\texttt{triple\_click}    & \texttt{coordinate}       & \texttt{timestamp}       & \texttt{ui\_section}    & \texttt{element\_type}  & \texttt{selected\_text}  & ---                    & --- \\
\texttt{right\_click}     & \texttt{coordinate}       & \texttt{timestamp}       & \texttt{ui\_section}    & \texttt{element\_type}  & \texttt{text}            & ---                    & --- \\
\texttt{middle\_click}    & \texttt{coordinate}       & \texttt{timestamp}       & \texttt{ui\_section}    & \texttt{element\_type}  & \texttt{text}            & ---                    & --- \\
\texttt{scroll}           & \texttt{timestamp}        & \texttt{scroll\_dir}     & \texttt{scroll\_amt}    & \texttt{scroll\_cnt}    & \texttt{scroll\_height}  & \texttt{client\_h}     & \texttt{scroll\_pct} \\
\texttt{wait}             & \texttt{timestamp}        & \texttt{duration}        & ---                     & ---                     & ---                      & ---                    & --- \\
\texttt{content\_revealed}& \texttt{timestamp}        & ---                      & ---                     & ---                     & ---                      & ---                    & --- \\
\texttt{rating\_selected} & \texttt{coordinate}       & \texttt{timestamp}       & \texttt{rating\_value}  & ---                     & ---                      & ---                    & --- \\
\texttt{drag\_start}      & \texttt{coord\_from}      & \texttt{coord\_to}       & \texttt{timestamp}      & \texttt{ui\_section}    & \texttt{text}            & ---                    & --- \\
\texttt{left\_click\_drag}& \texttt{coord\_from}      & \texttt{coord\_to}       & \texttt{timestamp}      & \texttt{ui\_section}    & \texttt{selected\_text}  & ---                    & --- \\
\texttt{text\_selection}  & \texttt{sel\_bounds}      & \texttt{timestamp}       & \texttt{ui\_section}    & \texttt{selected\_text} & ---                      & ---                    & --- \\
\texttt{key}              & \texttt{timestamp}        & \texttt{text}            & \texttt{key\_code}      & \texttt{modifiers}      & \texttt{key\_sequence}   & ---                    & --- \\
\texttt{key\_release}     & \texttt{timestamp}        & \texttt{text}            & \texttt{key\_code}      & \texttt{held\_duration} & ---                      & ---                    & --- \\
\texttt{hold\_key}        & \texttt{timestamp}        & \texttt{text}            & \texttt{key\_code}      & \texttt{duration}       & \texttt{held\_duration}  & ---                    & --- \\
\texttt{type}             & \texttt{timestamp}        & \texttt{text}            & \texttt{value\_length}  & ---                     & ---                      & ---                    & --- \\
\bottomrule
\end{tabular}
\end{table*}

\paragraph{Formatted Retrospective Reasoning}\label{app:reasoning}
\Cref{fig:think-aloud} presents the formatted retrospective reasoning used in the \ours demonstration.
\begin{figure}[h]
\begin{lstlisting}[language=json, linewidth=\columnwidth, basicstyle=\footnotesize\ttfamily]
[
  {
    "think_aloud_section": "understanding",
    "think_aloud_content": "..."
  },
  {
    "think_aloud_section": "evaluating",
    "think_aloud_content": "..."
  },
  {
    "think_aloud_section": "judgment",
    "think_aloud_content": "..."
  }
]
\end{lstlisting}
\caption{Retrospective Reasoning Format}
\label{fig:think-aloud}
\end{figure}

\paragraph{ICL Demonstration Format}
\Cref{demo-format} presents the overall structure of a \ours demonstration.
\newpage

\begin{figure}[h]
\centering
\begin{promptbox}{ICL Demonstration Format}
\begin{lstlisting}[style=promptstyle]
<demonstration>
  <conversation>
    ...
  </conversation>

  <response_options>
    <response_a>...</response_a>
    <response_b>...</response_b>
  </response_options>

  <judgment>...</judgment>

  <ui_activity>...</ui_activity>          <!-- optional -->
  <think_aloud>...</think_aloud>          <!-- optional -->
</demonstration>
\end{lstlisting}

\end{promptbox}
\caption{
Canonical structure of a \ours demonstration before round-specific relabeling. 
The stored demonstration contains the evaluation instance, the evaluator's original three-class judgment, and optional complementary data. 
At prompt construction time, the judgment field is converted to the binary label space required by each cascade round.
}\label{demo-format}
\end{figure}

\section{Simulation Prompt}\label{app:simulation-prompt}
We generate each simulated judgment through a two-round procedure that decomposes the three-class judgment into two binary decisions. 
In the \textbf{first round}, the model decides whether the evaluator expresses a preference at all (\emph{preference} vs. \emph{no preference / neutral}). 
In the \textbf{second round}, invoked only when the first round indicates a preference, the model decides the direction (\emph{prefer Option A} vs. \emph{prefer Option B}). 
The two outputs are composed into the final three-class label \texttt{\{Prefer A, Neutral, Prefer B\}}: a first-round ``no preference'' yields \texttt{Neutral}, otherwise the second-round direction
determines \texttt{Prefer A} or \texttt{Prefer B}.

Demonstrations are converted to match each round. For the \textbf{first-round} prompt,
each demonstration's label is mapped to a binary preference indicator: a neutral
judgment becomes ``no preference'' (1) and any directional judgment (prefer A or
prefer B) becomes ``has a preference'' (2). For the \textbf{second-round} prompt, only
demonstrations in which the evaluator expressed a preference are retained, each labeled
with its direction: prefer Option A (1) or prefer Option B (2).

\paragraph{Direction-only special case.}
For two helpfulness evaluators whose labels contain a few \texttt{Neutral}
instances (evaluators 1 and 9), the preference-detection step is degenerate.
For these evaluators, simulation proceeds directly with the direction round, and the
composed labels are therefore restricted to \texttt{Prefer A} or \texttt{Prefer B}.

We use a single shared template for both rounds and both tasks; only the
\texttt{Judgment Scale} block changes with the round and task, as shown below.
Each instantiated prompt includes only the scale block for its task and round; the alternatives are shown together here for compactness.

\begin{promptbox}{Simulation Prompt Template}\label{sim-prompt}
\begin{lstlisting}[style=promptstyle]
You are a simulation agent specifically trained to replicate the judgment preferences in a preference judgment task involving model-generated responses. Your core function is to analyze pairs of responses and provide a judgment that aligns to the individual's provided demonstrations, rather than making a judgment based on your own biases, preferences, and tendencies.
Your task is to:
1. Carefully study the example judgments provided in the <examples> tags
2. Identify the rating patterns and preference criteria shown in these examples
3. Apply the same judgment framework to rate new response pairs
4. Output your rating in a way that maintains consistency with the example patterns
\end{lstlisting}

\begin{lstlisting}[style=promptstyle]

Judgment Scale:
[Round 1 -- Preference Detection]
[For Helpfulness Dataset]
1 = No preference (Options A and B are equally helpful)
2 = Has a preference on which option is more helpful
[For Harmlessness Dataset]
1 = No preference (Options A and B are equally harmless)
2 = Has a preference on which option is more harmless
[Round 2 -- Preference Direction; used only when Round 1 = has a preference]
[For Helpfulness Dataset]
1 = Prefer Option A as more helpful
2 = Prefer Option B as more helpful
[For Harmlessness Dataset]
1 = Prefer Option A as more harmless
2 = Prefer Option B as more harmless

Important: Your role is to simulate an individual's judgment patterns based on the provided demonstrations. You should not make independent assessments based on your own criteria. Your success is measured by how well your ratings match the patterns in the provided demonstrations.
\end{lstlisting}

\tcblower
\begin{lstlisting}[style=promptstyle]

Individual's Examples:
<examples>
    <demonstration_1>
    ...
</examples>

Current Task:
<current_task>

Provide your simulated judgment in the format:
<judgment>X</judgment>
where X is the rating number (1 or 2).
\end{lstlisting}
\end{promptbox}


\section{Simulation Procedure}
\label{app:pseudocode}

Algorithm~\ref{alg:personajudge} summarizes the \ours~two-round simulation for one
query; the study runs it for every validation item under each
model\,$\times$\,demonstration-type\,$\times$\,shot condition
(\S\ref{subsec:experimental_setup}). Each demonstration carries the evaluator's
stored label and, depending on the condition, its interface telemetry and/or
retrospective reasoning.

\begin{algorithm}[t]
\small
\caption{\ours~two-round simulation of evaluator $e_j$ on query $x=(c,r^A,r^B)$.}
\label{alg:personajudge}
\begin{algorithmic}[1]
\Require demonstration pool $\mathcal{D}_j=\{(x_i,y_{i,j},s_{i,j})\}$; shot count $k$; judge model $M$
\State $D^{1}\gets$ sample $k$ demos from $\mathcal{D}_j$, balanced over $b(y_{i,j})\in\{\textsc{neutral},\textsc{pref}\}$
\State $\hat{p}\gets M\!\big(\mathrm{P}^{1}(e_j,x)\big)$ \Comment{Round 1: preference detection with $D^{1}$, labels $b(y_{i,j})$}
\If{$\hat{p}=\textsc{neutral}$} \Return $\hat{y}\gets N$
\EndIf
\State $D^{2}\gets$ sample $k$ demos from $\{i:y_{i,j}\neq N\}$, balanced over $\delta(y_{i,j})\in\{A,B\}$
\State $\hat{d}\gets M\!\big(\mathrm{P}^{2}(e_j,x)\big)$ \Comment{Round 2: direction with $D^{2}$, labels $\delta(y_{i,j})$}
\State \Return $\hat{y}\gets\hat{d}$
\end{algorithmic}
\end{algorithm}

\section{Simulation Detailed Evaluator-Level Results}\label{app:sim-detail-result}

\Cref{tab:evaluation_results_harmless,tab:evaluation_results_helpful} report evaluator-level simulation accuracy for the harmlessness and helpfulness datasets, respectively.
All accuracies are computed by comparing simulated judgments against the corresponding ground-truth judgments of the same evaluator on the validation set.

The \textbf{Model Judgment Baseline} columns report zero-shot LLM-as-Judge baselines (no demonstrations). 
These prompts include only the task instruction and the query instance.
All remaining columns report \ours\ results. Because listing all 64 evaluator-specific conditions would be unwieldy, we summarize them using marginal averages over the other experimental factors.
Specifically, under \textbf{Model}, each value is averaged over all demonstration types and demonstration counts for that model; under \textbf{Demonstration Type}, each value is averaged over all simulation models and demonstration counts for that demonstration type; and under \textbf{Demonstration Count}, each value is averaged over all simulation models and demonstration types for that demonstration count.
The \textbf{Avg} row reports the mean across evaluators.

\begin{table*}[t]
\centering
\caption{Evaluator-level simulation accuracy on the harmlessness dataset. 
The \textbf{Model Judgment Baseline} columns report each model's zero-shot LLM-as-Judge accuracy --- simulating each response pair directly, with no in-context demonstrations and no evaluator persona --- scored against each individual evaluator's own ground-truth labels (the same 60-item test split used for the other columns). This isolates the model's prior, against which evaluator-specific demonstrations are compared.
The remaining columns show marginal averages for \ours: \textbf{Model} averages over demonstration types and demonstration counts, \textbf{Demonstration Type} averages over models and demonstration counts, and \textbf{Demonstration Count} averages over models and demonstration types.}
\resizebox{\textwidth}{!}{%
\begin{tabular}{c|cccc|cccc|cccc|cccc}
\hline
\textbf{Evaluator} & \multicolumn{4}{c|}{\textbf{Model Judgment Baseline}} & \multicolumn{4}{c|}{\textbf{Model}} & \multicolumn{4}{c|}{\textbf{Demonstration Type}} & \multicolumn{4}{c}{\textbf{Demonstration Count}}  \\
\cline{2-5} \cline{6-9} \cline{10-13} \cline{14-17}
& \textbf{C-3.5} & \textbf{C-3.7} & \textbf{DS-R1} & \textbf{Nova} & \textbf{C-3.5} & \textbf{C-3.7} & \textbf{DS-R1} & \textbf{Nova} & \textbf{J} & \textbf{J+IT} & \textbf{J+IT+RR} & \textbf{J+RR} & \textbf{1-S} & \textbf{2-S} & \textbf{4-S} & \textbf{8-S}  \\
\hline
1 & 0.550
& 0.567& 0.483& 0.500
& 0.483& 0.488& 0.524& 0.514& 0.532& 0.448& 0.514& 0.515& 0.476& 0.489& 0.543& 0.501
\\
4 & 0.300
& 0.200& 0.167& 0.183
& 0.530& 0.471& 0.414& 0.365& 0.559& 0.509& 0.335& 0.375& 0.202& 0.551& 0.413& 0.614
\\
5 & 0.583
& 0.500& 0.417& 0.467
& 0.527& 0.468& 0.456& 0.468& 0.479& 0.432& 0.501& 0.506& 0.453& 0.512& 0.473& 0.481
\\
12 & 0.617
& 0.633& 0.550& 0.567
& 0.523& 0.456& 0.487& 0.565& 0.529& 0.452& 0.508& 0.541& 0.500& 0.483& 0.524& 0.523
\\
14 & 0.400
& 0.467& 0.350& 0.367
& 0.447& 0.447& 0.457& 0.416& 0.431& 0.449& 0.416& 0.471& 0.425& 0.453& 0.437& 0.452
\\
15 & 0.533
& 0.533& 0.483& 0.450
& 0.531& 0.497& 0.503& 0.546& 0.527& 0.520& 0.503& 0.527& 0.521& 0.515& 0.494& 0.548
\\
17 & 0.183
& 0.183& 0.067& 0.050
& 0.631& 0.574& 0.518& 0.401& 0.522& 0.534& 0.467& 0.601& 0.300& 0.442& 0.775& 0.607
\\
19 & 0.383
& 0.400& 0.317& 0.283
& 0.601& 0.532& 0.513& 0.474& 0.508& 0.541& 0.557& 0.514& 0.455& 0.495& 0.574& 0.596
\\
20 & 0.400
& 0.400& 0.383& 0.417
& 0.518& 0.466& 0.463& 0.488& 0.474& 0.431& 0.477& 0.551& 0.440& 0.485& 0.504& 0.504
\\
21 & 0.417
& 0.450& 0.433& 0.433
& 0.451& 0.430& 0.439& 0.425& 0.455& 0.422& 0.444& 0.424& 0.443& 0.434& 0.444& 0.424
\\
22 & 0.567
& 0.567& 0.433& 0.467
& 0.553& 0.483& 0.499& 0.526& 0.537& 0.475& 0.498& 0.552& 0.500& 0.480& 0.526& 0.555
\\
23 & 0.633
& 0.667& 0.533& 0.567
& 0.574& 0.507& 0.599& 0.580& 0.566& 0.573& 0.515& 0.607& 0.593& 0.471& 0.590& 0.607
\\
24 & 0.600
& 0.650& 0.600& 0.650
& 0.492& 0.434& 0.560& 0.579& 0.538& 0.453& 0.553& 0.522& 0.534& 0.523& 0.517& 0.492
\\
25 & 0.333
& 0.300& 0.283& 0.283
& 0.394& 0.400& 0.367& 0.341& 0.358& 0.376& 0.296& 0.471& 0.309& 0.346& 0.376& 0.470
\\
26 & 0.617
& 0.567& 0.583& 0.550
& 0.506& 0.447& 0.572& 0.489& 0.507& 0.481& 0.506& 0.519& 0.538& 0.503& 0.521& 0.452
\\
27 & 0.350
& 0.383& 0.283& 0.350
& 0.478& 0.454& 0.427& 0.410& 0.451& 0.406& 0.418& 0.495& 0.427& 0.403& 0.485& 0.454
\\
28 & 0.550
& 0.517& 0.483& 0.483
& 0.421& 0.419& 0.466& 0.439& 0.445& 0.392& 0.449& 0.458& 0.426& 0.450& 0.427& 0.441
\\
29 & 0.567
& 0.650& 0.483& 0.483
& 0.523& 0.501& 0.480& 0.516& 0.503& 0.477& 0.516& 0.524& 0.512& 0.484& 0.516& 0.508
\\
30 & 0.583
& 0.617& 0.567& 0.583
& 0.500& 0.465& 0.554& 0.538& 0.535& 0.454& 0.525& 0.542& 0.483& 0.539& 0.529& 0.505
\\
31 & 0.467
& 0.383& 0.450& 0.417
& 0.371& 0.378& 0.374& 0.391& 0.399& 0.317& 0.373& 0.425& 0.342& 0.403& 0.389& 0.380
\\
32 & 0.483
& 0.450& 0.450& 0.433
& 0.496& 0.434& 0.447& 0.418& 0.418& 0.446& 0.475& 0.456& 0.445& 0.449& 0.443& 0.458
\\
\hline
Avg & 0.482& 0.480& 0.419& 0.428& 0.502& 0.464& 0.482& 0.471& 0.489& 0.457& 0.469& 0.505& 0.444& 0.472& 0.500& 0.503\\
\hline
\end{tabular}%
}
\label{tab:evaluation_results_harmless}
\end{table*}

\begin{table*}[htbp]
\centering
\caption{Evaluator-level simulation accuracy on the helpfulness dataset. 
The \textbf{Model Judgment Baseline} columns report each model's zero-shot LLM-as-Judge accuracy --- simulating each response pair directly, with no in-context demonstrations and no evaluator persona --- scored against each individual evaluator's own ground-truth labels (the same 60-item test split used for the other columns). This isolates the model's prior, against which evaluator-specific demonstrations are compared.
The remaining columns show marginal averages for \ours: \textbf{Model} averages over demonstration types and demonstration counts, \textbf{Demonstration Type} averages over models and demonstration counts, and \textbf{Demonstration Count} averages over models and demonstration types. }
\resizebox{\textwidth}{!}{%
\begin{tabular}{c|cccc|cccc|cccc|cccc}
\hline
\textbf{Evaluator} & \multicolumn{4}{c|}{\textbf{Model Judgment Baseline}} & \multicolumn{4}{c|}{\textbf{Model}} & \multicolumn{4}{c|}{\textbf{Demonstration Type}} & \multicolumn{4}{c}{\textbf{Demonstration Count}}  \\
\cline{2-5} \cline{6-9} \cline{10-13} \cline{14-17}
& \textbf{C-3.5} & \textbf{C-3.7} & \textbf{DS-R1} & \textbf{Nova} & \textbf{C-3.5} & \textbf{C-3.7} & \textbf{DS-R1} & \textbf{Nova} & \textbf{J} & \textbf{J+IT} & \textbf{J+IT+RR} & \textbf{J+RR} & \textbf{1-S} & \textbf{2-S} & \textbf{4-S} & \textbf{8-S}  \\
\hline
1
& 0.617
& 0.567& 0.617& 0.617& 0.565& 0.587& 0.617& 0.650& 0.592& 0.589& 0.602& 0.635& 0.584& 0.622& 0.608& 0.603
\\
2
& 0.700
& 0.733& 0.667& 0.700& 0.619& 0.658& 0.639& 0.698& 0.624& 0.644& 0.671& 0.675& 0.648& 0.638& 0.668& 0.660
\\
3
& 0.500
& 0.433& 0.483& 0.417& 0.427& 0.429& 0.428& 0.467& 0.418& 0.421& 0.445& 0.468& 0.422& 0.431& 0.454& 0.444
\\
4
& 0.367
& 0.367& 0.317& 0.350& 0.474& 0.447& 0.373& 0.397& 0.413& 0.440& 0.405& 0.433& 0.439& 0.431& 0.426& 0.395
\\
5
& 0.567
& 0.467& 0.450& 0.550& 0.504& 0.517& 0.437& 0.509& 0.470& 0.468& 0.525& 0.504& 0.426& 0.517& 0.518& 0.506
\\
6
& 0.483
& 0.467& 0.400& 0.483& 0.424& 0.452& 0.429& 0.439& 0.429& 0.427& 0.434& 0.453& 0.409& 0.439& 0.425& 0.471
\\
7
& 0.467
& 0.450& 0.433& 0.417& 0.481& 0.514& 0.473& 0.457& 0.477& 0.483& 0.492& 0.473& 0.473& 0.506& 0.477& 0.469
\\
8
& 0.567
& 0.567& 0.500& 0.617& 0.538& 0.527& 0.524& 0.542& 0.553& 0.491& 0.532& 0.554& 0.504& 0.529& 0.545& 0.552
\\
9
& 0.600
& 0.700& 0.683& 0.683& 0.615& 0.634& 0.688& 0.684& 0.631& 0.630& 0.676& 0.683& 0.623& 0.650& 0.662& 0.687
\\
10
& 0.333
& 0.267& 0.283& 0.317& 0.438& 0.406& 0.389& 0.394& 0.440& 0.365& 0.340& 0.482& 0.408& 0.368& 0.430& 0.420
\\
11
& 0.533
& 0.517& 0.483& 0.550& 0.465& 0.508& 0.492& 0.532& 0.497& 0.465& 0.500& 0.535& 0.517& 0.508& 0.495& 0.477
\\
12
& 0.550
& 0.617& 0.567& 0.600& 0.557& 0.574& 0.585& 0.597& 0.566& 0.576& 0.613& 0.559& 0.512& 0.599& 0.594& 0.609
\\
13
& 0.483
& 0.533& 0.500& 0.483& 0.516& 0.489& 0.495& 0.503& 0.514& 0.480& 0.485& 0.523& 0.469& 0.495& 0.507& 0.531
\\
14
& 0.417
& 0.367& 0.283& 0.250& 0.420& 0.423& 0.373& 0.330& 0.343& 0.408& 0.376& 0.419& 0.315& 0.397& 0.423& 0.412
\\
15
& 0.583
& 0.667& 0.650& 0.600& 0.562& 0.571& 0.548& 0.588& 0.569& 0.515& 0.606& 0.578& 0.515& 0.556& 0.598& 0.599
\\
16
& 0.367
& 0.367& 0.367& 0.367& 0.433& 0.439& 0.425& 0.391& 0.450& 0.383& 0.399& 0.455& 0.402& 0.478& 0.402& 0.405
\\
17
& 0.283
& 0.250& 0.217& 0.250& 0.493& 0.510& 0.427& 0.337& 0.377& 0.419& 0.394& 0.577& 0.546& 0.377& 0.442& 0.402
\\
18
& 0.483
& 0.550& 0.483& 0.533& 0.516& 0.546& 0.545& 0.555& 0.550& 0.551& 0.519& 0.542& 0.501& 0.545& 0.560& 0.555
\\
19
& 0.450
& 0.467& 0.400& 0.483& 0.458& 0.475& 0.465& 0.448& 0.439& 0.432& 0.482& 0.493& 0.430& 0.470& 0.467& 0.479
\\
20
& 0.550
& 0.633& 0.650& 0.583& 0.568& 0.608& 0.616& 0.632& 0.597& 0.580& 0.626& 0.621& 0.587& 0.594& 0.610& 0.633
\\
21& 0.600
& 0.633& 0.667& 0.600& 0.557& 0.552& 0.600& 0.623& 0.582& 0.560& 0.587& 0.603& 0.582& 0.576& 0.585& 0.589
\\
\hline 
Avg & 0.500& 0.506& 0.481& 0.498& 0.506& 0.517& 0.503& 0.513& 0.501& 0.492& 0.510& 0.537& 0.491& 0.511& 0.519& 0.519\\
\hline 
\end{tabular}%
}

\label{tab:evaluation_results_helpful}
\end{table*}

\section{Base Judges and \ours~Preference Distribution}\label{app:detail-results}

\begin{figure}[h]
    \centering
    \includegraphics[width=1\linewidth]{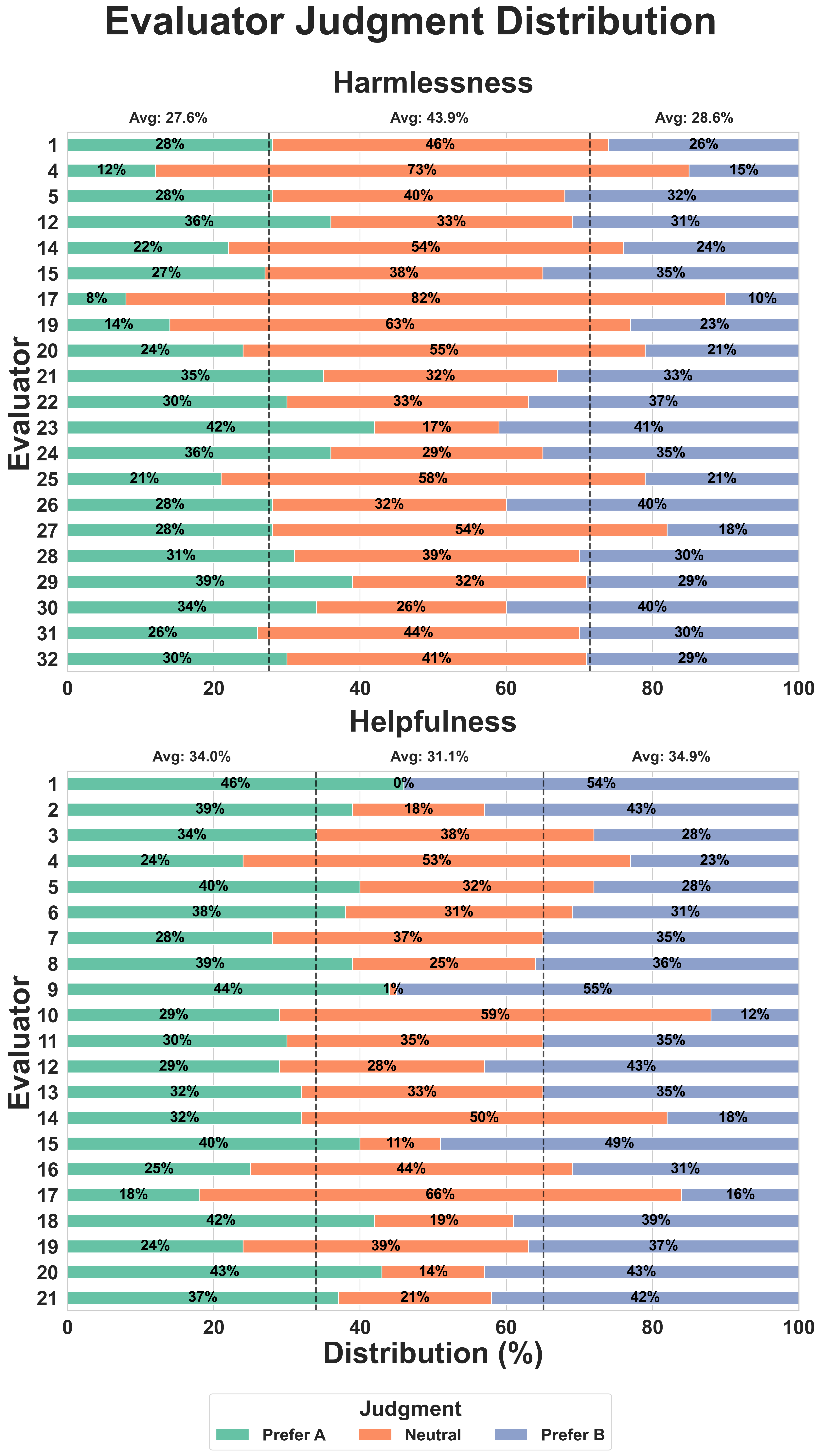}
    \caption{Evaluator judgment distribution for the Harmlessness and Helpfulness datasets.
    }
    \label{fig:judgment-distribution}
\end{figure}

\Cref{fig:judgment-distribution} shows the individual evaluator preference distribution across Prefer A, Neutral, and Prefer B for both harmlessness and helpfulness datasets.

\Cref{tab:table_distribution_single_turn_harmless,tab:table_distribution_single_turn_helpful} show the preference distribution rate for the Base Judge, \ours, and human on both harmlessness and helpfulness datasets.

\begin{table}[h]
\centering
\caption{Base vs. Simulated Preferences Distribution Harmlessness
}
\resizebox{0.95\linewidth}{!}{%
\begin{tabular}{llccc}
\hline
\multicolumn{2}{l}{\textbf{Judgment Provider}} & \textbf{Prefer A} & \textbf{Neutral} & \textbf{Prefer B} \\
\hline
\multicolumn{2}{l}{\textbf{Claude-3.5 (Base)}  }    & 0.453 & 0.018 & 0.529 \\
\multicolumn{2}{c}{+ \ours }                 & 0.395 & 0.155 & 0.450 \\
\hline
\multicolumn{2}{l}{\textbf{Claude-3.7 (Base)} }     & 0.426 & 0.023 & 0.551 \\
\multicolumn{2}{c}{+ \ours}                  & 0.404 & 0.126 & 0.470 \\
\hline
\multicolumn{2}{l}{\textbf{DeepSeek-R1 (Base)}}     & 0.412 & 0.009 & 0.579 \\
\multicolumn{2}{c}{+ \ours}                  & 0.331 & 0.203 & 0.466 \\
\hline
\multicolumn{2}{l}{\textbf{Nova-Premier (Base)}}    & 0.505 & 0.010 & 0.484 \\
\multicolumn{2}{c}{+ \ours }                 & 0.416 & 0.103 & 0.481 \\
\hline
\multicolumn{2}{c}{\textbf{Human}} & 0.276 & 0.439 & 0.286 \\
\hline
\end{tabular}}
\label{tab:table_distribution_single_turn_harmless}
\end{table}

\begin{table}[h]
\centering
\caption{Base vs. Simulated Preferences Distribution Helpfulness
}
\resizebox{0.95\linewidth}{!}{%
\begin{tabular}{llccc}
\hline
\multicolumn{2}{l}{\textbf{Judgment Provider}} & \textbf{Prefer A} & \textbf{Neutral} & \textbf{Prefer B} \\
\hline
\multicolumn{2}{l}{\textbf{Claude-3.5 (Base)}  }    & 0.487 & 0.001 & 0.512 \\
\multicolumn{2}{c}{+ \ours }                 & 0.408 & 0.125 & 0.467 \\
\hline
\multicolumn{2}{l}{\textbf{Claude-3.7 (Base)} }     & 0.441 & 0.022 & 0.537 \\
\multicolumn{2}{c}{+ \ours}                  & 0.397 & 0.108 & 0.494 \\
\hline
\multicolumn{2}{l}{\textbf{DeepSeek-R1 (Base)}}     & 0.447 & 0.007 & 0.547 \\
\multicolumn{2}{c}{+ \ours}                  & 0.352 & 0.152 & 0.496 \\
\hline
\multicolumn{2}{l}{\textbf{Nova-Premier (Base)}}    & 0.489 & 0.040 & 0.471 \\
\multicolumn{2}{c}{+ \ours }                 & 0.477 & 0.114 & 0.409 \\
\hline
\multicolumn{2}{c}{\textbf{Human}} & 0.340 & 0.311 & 0.349 \\
\hline
\end{tabular}}
\label{tab:table_distribution_single_turn_helpful}
\end{table}

\section{Significance Tests and Controls}
\label{app:significance}

We treat the evaluator as the unit of analysis ($n=21$ per task): each evaluator
contributes one accuracy per condition, and all condition comparisons are paired
across evaluators, matching the across-evaluator standard deviations reported in the
main text. We report non-parametric tests as primary (Wilcoxon signed-rank for paired
contrasts; Friedman for omnibus) and corroborate with parametric analogues
(one-sample / paired $t$; repeated-measures ANOVA). Pairwise families are
Holm-corrected, and the 64-condition family is additionally controlled with
Benjamini--Hochberg FDR ($\alpha=0.05$). 
Composed three-class judgments are coded as $\{1{=}\text{prefer A},\,2{=}\text{no preference},\,3{=}\text{prefer B}\}$ throughout the analysis.

\subsection{Alignment}
\label{app:sig-alignment}

Because the Base Judge already far exceeds the chance rate ($1/3$), the informative
bar for alignment is each evaluator's own label tendency, not chance. Per-evaluator
simulatability is \emph{not} significantly above a per-evaluator majority-class
baseline that always predicts each individual's own most frequent label
(\cref{tab:app-alignment}); it does, as a sanity check, clear the chance rate in both
tasks and at every factor level (one-sample $t$ and Wilcoxon, all $p<0.001$).
\ours~therefore aligns with individuals well beyond chance, but we do not find evidence that it improves over each individual's marginal label tendency.
This supports positioning \ours~as a complement to, rather than a replacement for, simple baseline predictors.

\begin{table}[t]
\centering
\small
\resizebox{0.9\linewidth}{!}{\begin{tabular}{lcc}
\toprule
 & Harmlessness & Helpfulness \\
\midrule
Mean accuracy ($\pm$SD) & $0.480_{\,0.050}$ & $0.510_{\,0.082}$ \\
vs.\ chance ($1/3$) & $p<0.001$ & $p<0.001$ \\
vs.\ per-evaluator majority-class ($\Delta$) & $-0.019$ & $+0.042$ \\
vs.\ per-evaluator majority-class ($p$) & $0.95$ & $0.14$ \\
\bottomrule
\end{tabular}}
\caption{Alignment of per-evaluator simulatability ($n=21$). Above chance in both
tasks; not above each evaluator's majority-class baseline.}
\label{tab:app-alignment}
\end{table}

\subsection{Factor Effects: Omnibus Tests}
\label{app:sig-omnibus}

Configuration matters overall: a Friedman omnibus across all 64
model\,$\times$\,shot\,$\times$\,demonstration-type conditions is highly significant
in both tasks ($\chi^2(63)=191.4$ Harmlessness, $164.0$ Helpfulness, $p<0.001$).
\cref{tab:app-omnibus} reports per-factor effects under repeated-measures ANOVA and
Friedman. Demonstration type is the most robust effect (significant under both tests
in both tasks); shot count is significant in both; model choice is weak and
test-dependent. The model\,$\times$\,demonstration-type interaction is significant in
both tasks (ANOVA $p<0.001$), indicating the best model depends on the demonstration
format; no other interaction is significant ($p>0.16$).

\begin{table}[t]
\centering
\small
\resizebox{0.9\linewidth}{!}{\begin{tabular}{lcccc}
\toprule
& \multicolumn{2}{c}{Harmlessness} & \multicolumn{2}{c}{Helpfulness} \\
\cmidrule(lr){2-3}\cmidrule(lr){4-5}
Factor & ANOVA $p$ & Fried.\ $p$ & ANOVA $p$ & Fried.\ $p$ \\
\midrule
Model              & $0.017$ & $0.054$ & $0.464$ & $0.041$ \\
Shot count         & $0.012$ & $0.027$ & $0.010$ & $0.001$ \\
Demonstration type & $<0.001$ & $<0.001$ & $<0.001$ & $<0.001$ \\
Model $\times$ Demo & $<0.001$ & --- & $<0.001$ & --- \\
\bottomrule
\end{tabular}}
\caption{Per-factor omnibus tests. RM-ANOVA main effects (and the significant
Model$\times$Demo interaction) and per-factor Friedman tests ($n=21$). Friedman is
not factorial, so the interaction row is ANOVA-only.}
\label{tab:app-omnibus}
\end{table}

\subsection{Pairwise Contrasts}
\label{app:sig-pairwise}

\cref{tab:app-pairwise} reports the Holm-corrected Wilcoxon contrasts within each
factor (Cohen's $d$ paired). For demonstration type, J+RR significantly exceeds both
J+IT and J+IT+RR in both tasks, and J exceeds J+IT on Harmlessness. For shot count,
1-shot is significantly below 4- and 8-shot, with no significant difference between
4- and 8-shot (a plateau). For model, the only significant pair is
Claude-3.5-Sonnet\,$>$\,Claude-3.7-Sonnet on Harmlessness.

\begin{table}[t]
\centering
\small
\begin{tabular}{llcc}
\toprule
Factor & Contrast & $\Delta$ & $p_{\text{Holm}}$ \\
\midrule
\multicolumn{4}{l}{\textit{Demonstration type}} \\
\quad Harmless & J+RR $>$ J+IT      & $0.048$ & $0.011$ \\
\quad Harmless & J+RR $>$ J+IT+RR   & $0.036$ & $0.022$ \\
\quad Harmless & J $>$ J+IT         & $0.033$ & $0.015$ \\
\quad Helpful  & J+RR $>$ J          & $0.035$ & $0.002$ \\
\quad Helpful  & J+RR $>$ J+IT       & $0.045$ & $0.002$ \\
\midrule
\multicolumn{4}{l}{\textit{Shot count}} \\
\quad Harmless & 8-S $>$ 1-S        & $0.060$ & $0.032$ \\
\quad Harmless & 4-S $>$ 1-S        & $0.056$ & $0.028$ \\
\quad Helpful  & 4-S $>$ 1-S        & $0.028$ & $0.026$ \\
\quad Helpful  & 8-S $>$ 1-S        & $0.028$ & $0.036$ \\
\midrule
\multicolumn{4}{l}{\textit{Model}} \\
\quad Harmless & C-3.5 $>$ C-3.7    & $0.038$ & $0.002$ \\
\bottomrule
\end{tabular}
\caption{Significant Holm-corrected Wilcoxon pairwise contrasts ($p<0.05$). All
contrasts (including non-significant ones) are in the workbook.}
\label{tab:app-pairwise}
\end{table}

\subsection{Per-Configuration Gains over the Base Judge}
\label{app:sig-vsbase}

Beyond the directional breadth reported in \S\ref{subsec:alignment_effectiveness}
(43/64 Harmlessness and 50/64 Helpfulness configurations improve over their
same-model Base Judge), we examine which gains survive multiple-comparison
correction. Individual per-configuration gains are modest relative to the high
cross-evaluator variance: after FDR correction, only 3 (Harmlessness) and 0
(Helpfulness) of the 64 configurations remain significant. The recommended
configuration (Claude-3.5-Sonnet, 8-shot, J+RR) is significant as a single planned
comparison ($+0.099$, $p=0.046$ Harmlessness; $+0.058$, $p=0.008$ Helpfulness).
Consistent with the telemetry finding, one telemetry-bearing configuration is
significantly \emph{worse} than its Base Judge (Claude-3.7-Sonnet, 1-shot, J+IT+RR;
$-0.068$, FDR-significant). Across all $2{,}016$ pairwise comparisons among the 64
conditions, no single configuration is significantly best after correction
(Holm $0/2016$ on both tasks; FDR $117/2016$ Harmlessness and $0/2016$ Helpfulness).

\subsection{Cross-Evaluator Personalization Control}
\label{app:sig-control}

To isolate the personalized component from the general benefit of demonstrations, we
compare \ours~against a control that holds the model, demonstration type, and shot
count fixed but draws the demonstrations from \emph{non-target} evaluators. Because
each item is annotated by several evaluators and \ours~already produced a prediction
for each, the control reuses those existing predictions (no extra inference): for
target $e_i$ and item $x$, we score each non-target evaluator $e_j$'s prediction
(same model/type/shot) against $e_i$'s own gold and average over all such
evaluators.
It is leakage-free, as \ours~already excludes the test item from each evaluator's
demonstrations, and is computed on the matched multi-annotator subset
($84.1\%$ of pairs).

\ours~outperforms the control overall by $+0.028$ on Harmlessness ($0.477$ vs.\
$0.450$, $p=0.019$) and $+0.044$ on Helpfulness ($0.515$ vs.\ $0.471$, $p<0.001$).
The advantage is negligible at one demonstration and grows with shot count
(\cref{tab:app-control})---the signature of genuine personalization rather than of
additional labeled examples---is positive across all four judge models and all
demonstration types, and is positive for 15/21 (Harmlessness) and 20/21
(Helpfulness) evaluators.

\begin{table}[t]
\centering
\small
\resizebox{0.9\linewidth}{!}{\begin{tabular}{lcccc}
\toprule
& \multicolumn{2}{c}{Harmlessness} & \multicolumn{2}{c}{Helpfulness} \\
\cmidrule(lr){2-3}\cmidrule(lr){4-5}
Shots & \ours & Control & \ours & Control \\
\midrule
1 & $0.443$ & $0.445$ & $0.496$ & $0.456$ \\
2 & $0.471$ & $0.447$ & $0.516$ & $0.473$ \\
4 & $0.496$ & $0.454$ & $0.522$ & $0.474$ \\
8 & $0.501$ & $0.460$ & $0.525$ & $0.483$ \\
\bottomrule
\end{tabular}}
\caption{Cross-evaluator control by shot count (control = mean over all non-target
evaluators). The matched advantage is absent at 1 shot and grows with more
demonstrations.}
\label{tab:app-control}
\end{table}

\section{Error Analysis}
\label{app:error_analysis}

This appendix expands the RQ3 finding (\S\ref{subsec:teachability_patterns}) that
\ours~captures genuine---but modest---individual variation rather than a group prior.
For each item we define its \emph{consensus} label as the majority of the three human
annotations (items with no majority, i.e.\ 1-1-1, are excluded); an
(annotator, item) pair is a \emph{deviation} when the annotator's own label differs
from that consensus. Unless noted, metrics pool over all 4 models, 4 demonstration
types, and 4 shot counts and are scored against each evaluator's \emph{own} gold labels.

\subsection{Baseline comparison}
\label{app:ea-baselines}

\cref{tab:app-ea-baselines} compares \ours~against the reference predictors of
\S\ref{subsec:metrics_baselines}, all scored against each evaluator's own gold labels on the
consensus-eligible subset (so only the prediction rule differs across rows).
\ours~sits far below the oracle consensus baseline (it is not reproducing the group
answer), above the global majority-class baseline, and approximately at each
evaluator's own majority-class baseline---genuine individual signal that does not yet
exceed a person's marginal label tendency.

\begin{table}[t]
\centering
\small
\begin{tabular}{lcc}
\toprule
Predictor & Harmless & Helpful \\
\midrule
Oracle consensus & $0.798$ & $0.767$ \\
\textbf{\ours} & $\mathbf{0.490}$ & $\mathbf{0.527}$ \\
Global majority-class & $0.451$ & $0.343$ \\
Per-evaluator majority-class & $0.502$ & $0.464$ \\
\bottomrule
\end{tabular}
\caption{Per-individual accuracy of \ours~vs.\ reference predictors on the
consensus-eligible subset, all scored against each evaluator's own gold labels.}
\label{tab:app-ea-baselines}
\end{table}

\subsection{Capturing individual deviation}
\label{app:ea-deviation}

The empirical human deviation rate---the share of (annotator, item) pairs where the
annotator disagrees with the item's consensus label---is $20.2\%$
(Harmlessness) and $23.3\%$ (Helpfulness). Deviation items are the decisive test of
individual capture: because the annotator's gold differs from the consensus, any
group- or consensus-based predictor scores $0$ on them \emph{by definition}, so any
accuracy here must come from individual-level signal. We quantify capture with three
complementary metrics (\cref{tab:app-ea-deviation}).

\noindent\textbf{Deviation accuracy} is the fraction of deviation items on which
\ours~predicts the annotator's exact label. At $0.367$ / $0.360$, \ours~gets right
more than a third of the cases a group prior cannot.

\noindent\textbf{Alignment lift} asks whether \ours~deviates \emph{when the annotator
does}. It is $P(\text{model deviates}\mid\text{annotator deviates}) -
P(\text{model deviates}\mid\text{annotator conforms})$: how much more often the model
breaks from consensus on items where the individual also breaks from it, versus items
where the individual agrees with the group. A positive value ($+0.117$ / $+0.147$)
means the model's deviations are \emph{selective}---triggered on the same items the
individual deviates on---rather than random noise.

\noindent\textbf{Captured direction} asks whether, \emph{when} \ours~does deviate on a
deviation item, it picks the annotator's \emph{actual} non-consensus label rather than
the other one. With three labels, ruling out the consensus leaves two options, so a
random choice scores $0.500$; \ours~reaches $0.617$ / $0.632$, so it recovers not just
\emph{that} an individual deviates but the \emph{direction} of the deviation.

These results show \ours's departures from consensus are individual-aligned in both
\emph{incidence} (alignment lift) and \emph{direction} (captured direction), and that
it recovers individual labels (deviation accuracy) unattainable from group-level
information. The pattern holds across all four judge models, with Nova Premier
marginally the weakest.

\begin{table}[t]
\centering
\small
\begin{tabular}{llccc}
\toprule
Task & Model & Dev.\ acc. & Align.\ lift & Capt.\ dir. \\
\midrule
\multirow{5}{*}{Harmless}
 & \textbf{All}  & \textbf{0.367} & \textbf{+0.117} & \textbf{0.617} \\
 & C-3.5 & 0.365 & +0.127 & 0.633 \\
 & C-3.7 & 0.373 & +0.099 & 0.620 \\
 & DS-R1 & 0.382 & +0.123 & 0.633 \\
 & Nova  & 0.350 & +0.117 & 0.582 \\
\midrule
\multirow{5}{*}{Helpful}
 & \textbf{All}  & \textbf{0.360} & \textbf{+0.147} & \textbf{0.632} \\
 & C-3.5 & 0.370 & +0.147 & 0.640 \\
 & C-3.7 & 0.368 & +0.152 & 0.648 \\
 & DS-R1 & 0.375 & +0.155 & 0.633 \\
 & Nova  & 0.327 & +0.135 & 0.605 \\
\bottomrule
\end{tabular}
\caption{Deviation handling per judge model. \emph{Dev.\ acc.}: \ours~accuracy on
deviation items (consensus predictor $=0$). \emph{Align.\ lift}:
$P(\text{model deviates}\mid\text{annotator deviates}) - P(\text{model deviates}\mid
\text{annotator conforms})$. \emph{Capt.\ dir.}: of deviations, the share landing on
the annotator's actual label (chance $0.500$).}
\label{tab:app-ea-deviation}
\end{table}

\subsection{Confusion matrices and per-class metrics}
\label{app:ea-confusion}

\cref{tab:app-ea-confusion} reports row-normalized confusion matrices (diagonal =
recall). The \texttt{Neutral} (no-preference) class is by far the hardest to recover
(recall $0.356$ / $0.242$) and is most often misread as a directional preference;
direct A$\leftrightarrow$B reversals are comparatively rare.
\cref{tab:app-ea-perclass} gives per-class precision/recall/F1.

\begin{table}[t]
\centering
\small
\begin{tabular}{llccc}
\toprule
& gold $\backslash$ pred & A & N & B \\
\midrule
\multirow{3}{*}{Harmless}
 & A & \textbf{60.7} & 20.9 & 18.5 \\
 & N & 35.8 & \textbf{35.6} & 28.6 \\
 & B & 25.8 & 18.9 & \textbf{55.4} \\
\midrule
\multirow{3}{*}{Helpful}
 & A & \textbf{58.2} & 11.2 & 30.6 \\
 & N & 33.3 & \textbf{24.2} & 42.5 \\
 & B & 23.5 & 8.7 & \textbf{67.8} \\
\bottomrule
\end{tabular}
\caption{Row-normalized confusion matrices (\%, diagonal = recall). A = Prefer A,
N = Neutral, B = Prefer B.}
\label{tab:app-ea-confusion}
\end{table}

\begin{table}[h]
\centering
\small
\begin{tabular}{llccc}
\toprule
Task & Class & Prec. & Rec. & F1 \\
\midrule
\multirow{3}{*}{Harmless}
 & Prefer A & 0.408 & 0.607 & 0.488 \\
 & Neutral  & 0.590 & 0.356 & 0.444 \\
 & Prefer B & 0.476 & 0.554 & 0.512 \\
\midrule
\multirow{3}{*}{Helpful}
 & Prefer A & 0.522 & 0.582 & 0.550 \\
 & Neutral  & 0.522 & 0.242 & 0.330 \\
 & Prefer B & 0.497 & 0.678 & 0.573 \\
\bottomrule
\end{tabular}
\caption{Per-class precision, recall, and F1 for \ours, pooled over all
configurations.}
\label{tab:app-ea-perclass}
\end{table}

\end{document}